\documentclass[3p,times]{elsarticle}
\usepackage{ecrc}
\volume{00}
\firstpage{1}
\journalname{Journal of Parallel and Distributed Computing}
\runauth{B. Guindani et al.}  
\jid{jdpc}
\usepackage[]{algorithm2e}
\RestyleAlgo{boxruled}
\LinesNumbered
\usepackage{amssymb}
\usepackage{amsmath}
\usepackage{booktabs} 
\usepackage{enumitem}
\usepackage{hyperref}  
\usepackage{subcaption}
\usepackage{xcolor}

\graphicspath{{images/}}
\DeclareMathOperator*{\argmin}{arg\,min}
\DeclareMathOperator*{\argmax}{arg\,max}

\begin{document}
\begin{frontmatter}

\title{Efficient Parameter Tuning for a Structure-Based Virtual Screening HPC Application}

\author[polimi]{Bruno Guindani}
\author[polimi]{Davide Gadioli}
\author[polimi]{Roberto Rocco}
\author[polimi]{Danilo Ardagna}
\author[polimi]{Gianluca Palermo}
\address[polimi]{Politecnico di Milano, Dipartimento di Elettronica, Informazione e Bioingegneria, via Giuseppe Ponzio 34/5, Milano}

\begin{abstract}
Virtual screening applications are highly parameterized to optimize the balance between quality and execution performance.
While output quality is critical, the entire screening process must be completed within a reasonable time.
In fact, a slight reduction in output accuracy may be acceptable when dealing with large datasets.
Finding the optimal quality-throughput trade-off depends on the specific HPC system used and should be re-evaluated with each new deployment or significant code update.
This paper presents two parallel autotuning techniques for constrained optimization in distributed High-Performance Computing (HPC) environments.
These techniques extend sequential Bayesian Optimization (BO) with two parallel asynchronous approaches, and they integrate predictions from Machine Learning (ML) models to help comply with constraints.
Our target application is LiGen, a real-world virtual screening software for drug discovery.
The proposed methods address two relevant challenges: efficient exploration of the parameter space and performance measurement using domain-specific metrics and procedures.
We conduct an experimental campaign comparing the two methods with a popular state-of-the-art autotuner.
Results show that our methods find configurations that are, on average, up to 35--42\% better than the ones found by the autotuner and the default expert-picked LiGen configuration.
\end{abstract}

\begin{keyword}
Application Autotuning \sep High-Performance Computing \sep Virtual Screening \sep Bayesian Optimization

\end{keyword}

\end{frontmatter}


\section{Introduction}
Drug discovery is a lengthy and costly process that aims to find a molecule that yields a therapeutic effect in clinical trials.
Traditionally, we start by measuring in-vitro the interaction strength between a ligand, i.e., the drug candidate, and the target protein(s) that represent the disease.
One problem is that, nowadays, we need to explore a chemical space in the order of $10^{33}$ ligands \cite{polishchuk2013estimation}.
Recent studies demonstrated that introducing a virtual screening stage to select the most promising ligands for in-vitro tests increases the drug discovery success probability \cite{matter2011application, allegretti2022repurposing}.
To estimate if a ligand is a good candidate, we need to solve two well-known problems in literature: molecular docking and scoring \cite{murugan2022review, pagadala2017software, su2018comparative}.
Both are complex tasks with no optimal solutions but rather a trade-off spectrum between throughput and accuracy.
Since we can evaluate each protein-ligand independently, this problem is embarrassingly parallel.
For these reasons, a High-Performance Computing (HPC) system is the ideal infrastructure for a virtual screening campaign.

If we look at the top ten supercomputers listed in the current version (as of June 2024) of the Top500 list\footnote{\url{https://www.top500.org}}, we can notice how much the hardware landscape is diverse but shares some similarities.
All but one of these supercomputers rely heavily on accelerators to increase their performance.
Even if most of them use GPUs, they belong to different generations and vendors.
This heterogeneity makes it impossible to define a unique, efficient configuration regardless of the underlying hardware.
Thus, we must adapt the application to fit the execution environment \cite{9309041}.
A virtual screening campaign might benefit from any speedup by evaluating more candidates, increasing the probability of finding good candidates.

A typical solution to cope with heterogeneous environments is to expose software knobs.
This way, we can tune them to tailor the application to the underlying architecture.
Moreover, scientific applications might employ heuristics and optimization methods that can expose knobs that impact the quality of the results as well \cite{gadioli2021tunable}.
Usually, the relationship between input features, performance, and quality of results is not trivial.
The automated tuning of these parameters is becoming a good practice in large-scale infrastructure \cite{verma2015large}, as also highlighted by the growing interest in this topic within the literature \cite{gocht2019q, balaprakash2018autotuning, rasch2018atf, dorn2017automatically}.

This work focuses on a real-world virtual screening application to show how we can perform autotuning at deployment time.
We address the following two challenges: \textit{i)} how to explore the parameter space in a supercomputer efficiently, and \textit{ii)} how to measure the performance using domain-specific metrics and procedures.
To solve the first challenge, we use enhanced Bayesian Optimization (BO) \cite{frazier2018tutorial} techniques to drive the selection of parameter configurations to explore.
Since, in the HPC context, we have access to a large number of nodes to spread the exploration, we develop different techniques that represent parallel versions of a sequential BO algorithm.
We implement these strategies by extending our framework MALIBOO \cite{guindani2024integrating} to a parallel setting.
Moreover, we also consider user requirements to constrain the parameter exploration within the region of interest for the user.
To orchestrate the autotuning phase on multiple nodes, we hinge on the HyperQueue scheduler \cite{Beranek2024HQ} for dispatching the evaluation of each configuration in parallel, lowering the exploration wall-clock time.
To address the second challenge (performance measurement), we define application performance as a function of execution time given a fixed input dataset and a domain-specific metric to assess the quality of results.
This solution is non-trivial as we cannot measure both metrics with a single run of the application, but we need to introduce an additional layer to evaluate a configuration using all the resources of a single node.
%


In this paper, we targeted LiGen, the virtual screening application of the EXSCALATE platform \cite{9817028}.
We define a metric that takes both execution time and solution quality into account and evaluate the proposed approach according to such metric.
We also compare the proposed approaches with OpenTuner, a state-of-the-art autotuner \cite{ansel2014opentuner}.
In particular, the contributions of this paper are the following:
\begin{itemize}
    \item We analyzed the target application characterized by the need for tuning for two reasons: performance-portability across different HPC platforms and throughput requirements that can vary if targeting small- or large-scale virtual screening campaigns;
    \item We explore two strategies that use BO techniques and ML models in a parallel environment: the ensemble-based EMaliboo and the centralized PAMaliboo techniques;
    \item We designed an exploration procedure that makes use of a meta-scheduler to hinge on the resources available in a supercomputer efficiently;
    \item We evaluate the benefits of the proposed autotuning approach at deployment time using a relevant case study.
\end{itemize}
From the experimental results, we observe how the proposed strategies improve state-of-the-art and a LiGen configuration chosen by experts.
Both parallel strategies are best suited for specific scenarios: EMaliboo has better exploration capabilities when few data are sufficient to build accurate surrogate models, whereas PAMaliboo provides superior robustness for larger datasets.

The remainder of the paper is structured as follows.
Section \ref{sec:bo-background} provides some background about sequential and parallel BO techniques.
Section \ref{sec:ligen} describes the target virtual screening application, giving the context required to state the parameter space and the metrics of interest; we also highlight why leveraging computation resources efficiently is paramount.
Section \ref{sec:autotuning} dives into the formalization of our optimization problem and our proposed autotuning algorithms, which we evaluate in Section \ref{sec:exper}.
Section \ref{sec:related} summarizes the state of the art related to application autotuners.
Finally, Section \ref{sec:conclusion} concludes the paper.

\section{Bayesian Optimization background}  \label{sec:bo-background}
A relevant branch of the software optimization literature exploits the Bayesian Optimization (BO) technique and its variations \cite{egele2022asynchronous, frisby2021asynchronous, kandasamy2018parallelised, snoek2012practical}.
BO has recently gained notoriety as a powerful tool for solving global optimization problems involving expensive, black-box functions.
These are functions for which little information is available and whose evaluation has significant time, resource, or monetary costs.
BO is a sequential design strategy that only needs a few steps to get sufficiently close to the true optimum and requires no strong assumptions or derivative information on the analyzed function.
For these reasons, the field considers BO a particularly efficient method to minimize this kind of objective function \cite{frazier2018tutorial, schonlau1998global}.

BO applies to both unconstrained and constrained optimization scenarios.
In particular, in a constrained global optimization setting, BO considers the following problem formulation:
\begin{align}
    \begin{split}
    \min_{x \in A} f(x) \\
    \text{s.t. } \quad g(x) &\in [G_{min}, G_{max}]
    \end{split}
    \label{eq:problem-general}
\end{align}
$x \in A$ denotes the $d$-dimensional vector representing a configuration for the system at hand, with $A \subset \mathbb{R}^d$ being the domain of all possible configurations.
We will interchangeably refer to such vectors as configurations or points.
$x$ includes the values of a (possibly large) number of software and hardware configuration parameters.
The black-box \textit{objective function} $f(x): A \to \mathbb{R}$ to minimize typically measures the performance or quality of configuration $x$ (e.g., a cross-validation error score), the running time or cost of an application, and so forth.
We also assume a scalar \textit{constraint function} $g(x): A \to \mathbb{R}$, which is possibly independent of the objective function.
The constraint $g(\cdot)$ is also a black-box function whose expression is unknown.
Therefore, the feasible part of the domain, i.e., the one containing points that satisfy the constraint, is also unknown.
We represent such constraint on $x$ with a potentially unbounded closed interval, whose extrema $G_{min}$ and $G_{max}$ can be infinite.
This formulation may address the applications that employ a stochastic method or heuristic to solve a complex task in a feasible time frame.
For example, the function $f(\cdot)$ might represent the throughput of the application, while the function $g(\cdot)$ represents the quality of results.
In BO settings, the \textit{Gaussian process} (GP) \cite{rasmussen2006gaussian} is the preferred choice for the prior distribution, or surrogate model, for $f(\cdot)$.
For any $x \in A$, this prior assigns to each value of $f(x)$ a Gaussian probability distribution which depends on $x$:
\begin{equation} \label{eq:prior}
    f(x) \sim \pi_x(\cdot) = \mathcal{N} (\mu_0(x), \sigma_0^2(x)).
\end{equation}
In Equation~\eqref{eq:prior}, $\mu_0(\cdot)$ and $\sigma_0^2(\cdot)$ denote the mean and kernel functions, respectively, and they constitute the GP model hyperparameters.
These functions serve as  ``initial guesses'' on values of $f(\cdot)$ and its uncertainty, a starting point for the BO algorithm, which will update them with observed values.
In particular, let $H_n = \{(x_1, f(x_1)), \dots, (x_n, f(x_n))\}$ be the history of $n$ past observations.
Precisely, observation $i$ consists of the configuration vector $x_i$ and the associated evaluation of the objective function $f(x_i)$.
Having observed values in $H$, one can compute the posterior distribution of each $f(x)$, starting from the prior distribution in Equation~\eqref{eq:prior} and considering these observations.
In the case of the GP prior, this means computing the updated hyperparameters $\mu_n(\cdot)$ (posterior mean) and $\sigma_n^2(\cdot)$ (posterior variance).
We then write the posterior distribution as $f(x) | H_n \sim \pi_x(\cdot | H_n) = \mathcal{N} (\mu_n(x), \sigma^2_n(x))$ (see \cite{frazier2018tutorial} for further details).

BO formulates a proxy problem at each step -- the maximization of the \textit{acquisition function}.
This function depends on the history and the GP model at the current algorithm iteration $n$ and measures the utility of evaluating the objective function $f(x)$ in a given configuration $x$.
Formally, we denote this as $a(x | H_n) : A \to \mathbb{R}$, or $a(x)$ for short.
The BO iterative algorithm optimizes this function at each round instead of directly optimizing the objective function itself.
Specifically, at step $n$, BO computes the maximizer $x_{n+1} = \argmax_{x \in A}{a(x | H_n)}$ of the current acquisition function (i.e., given the history $H_n$).
This point represents the next evaluation of the objective function, i.e., we compute $f(x_{n+1})$ and add it to the history of evaluated points.

\subsection{Synchronous vs asynchronous BO}  \label{subsec:backgr-sync-async}
Since classical BO is a sequential algorithm, it only performs one evaluation of the objective function at a time.
In settings with enough available computational power to allow multiple simultaneous evaluations, such as HPC systems, the sequential nature of BO represents a hindrance to efficiency.
Therefore, the literature has explored multiple attempts at extending the BO algorithm to a parallel setting \cite{egele2022asynchronous, frisby2021asynchronous, kandasamy2018parallelised, snoek2012practical, ginsbourger2010kriging}.

There are two main approaches to parallel BO optimization techniques, representing extremes of a spectrum: synchronous and asynchronous \cite{assran2020advances}.
Synchronous approaches choose one batch of points at each iteration in such a way that they jointly maximize a specific acquisition function \cite{zhang2020efficient}.
These approaches do not necessarily choose individual points that are optimal on their own but rather a set of points that are optimal in conjunction with each other, e.g., because they allow the best exploration possible given the current information.
Evaluation of these points happens at the same time, and the update to the BO model takes place afterward.
Other than the parallel evaluation of the objective function, pure synchronous algorithms are otherwise similar to non-parallel sequential approaches \cite{frazier2018tutorial}.

Synchronous approaches benefit from iteration efficiency, as they select each batch to balance exploration and exploitation jointly.
However, their drawback is that they still do not fully harness the full parallel capabilities of the system they are running on.
The only part of pure synchronous algorithms that exploit full parallelism is the simultaneous evaluation of the batch points, and they operate as sequential algorithms in all other respects.
This limitation is especially relevant in cases where there is substantial variance in the computation time of the objective function as the input configuration varies.
In these cases, the algorithm must wait a significant amount of time for the slowest evaluation to proceed to the next iteration.

Conversely, in asynchronous parallel approaches, objective evaluations are non-blocking for algorithm iterations \cite{zhang2020efficient}.
The algorithm requests the evaluation of one or more configurations whenever idle parallel workers are available and later integrates information from completed evaluations.
The update of the BO model takes place as soon as new information is available, i.e., when any objective evaluation has terminated, without waiting for the evaluations that are still running.
In this way, the asynchronous algorithm fully exploits the parallelism of the system at all times.
However, it might suffer from iteration inefficiency since it does not choose subsequent points jointly -- effectively being a one-step-ahead algorithm compared to the multiple-step-ahead synchronous ones \cite{ginsbourger2010kriging}.

Besides pure synchronous and asynchronous approaches, hybrid algorithms also exist.
For instance, an algorithm may simultaneously evaluate a batch of points whose size depends on the current available computational power.
In this case, the batch size represents a trade-off between the iteration efficiency of synchronous approaches (large batch size) and the computing resource efficiency of asynchronous approaches (small or unitary batch size).

This work focuses on asynchronous approaches since the available computational power is remarkably costly in HPC settings.
Therefore, we prioritize exploiting as much parallel power as possible within time and budget constraints rather than aiming for optimal iteration efficiency.

\subsection{Parallel BO algorithms}  \label{subsec:backgr-bo-algos}
There are many different possible approaches to parallel BO.
However, all approaches, whether synchronous, asynchronous, or hybrid, must cope with a problem arising from the deterministic nature of BO -- what we might call the \textit{quicksand problem}.  
A parallel BO algorithm must immediately choose one configuration for each parallel worker if two or more idle (out of $q$).
However, a fixed history of $n$ points always yields the same posterior distribution and acquisition function and, therefore, the same next point to evaluate, $x_{n+1}$.
Hence, it is impossible to sequentially choose different points from the same posterior distribution in a deterministic way.
Without new information to integrate into the posterior, the algorithm would be stuck on the same point, similar to being stuck in quicksand due to not applying enough pressure to escape.
Different techniques are characterized by how they are affected by this problem and how they attempt to solve it.

Pure synchronous approaches handle the quicksand problem within each iteration when choosing a batch of points from the same posterior distribution.
They employ multi-input acquisition functions that they maximize to obtain a jointly optimal batch of $q$ points.
Also, as previously mentioned, pure synchronous methods can hybridize with asynchronous approaches by employing an adaptive batch size equal to the number of currently idle workers while keeping a synchronous acquisition function that deals with the quicksand problem within each batch.
For instance, the well-known Expected Improvement (EI) acquisition function can extend to q-EI or multipoint EI \cite{frazier2018tutorial, ginsbourger2007multi}.
A similar approach is PROTOCOL \cite{frisby2021asynchronous}, a parallel extension of the BO-based IMGPO algorithm \cite{kawaguchi2015bayesian}.
IMGPO uses hierarchical domain partitioning to select points for BO to sample without needing a proxy optimization problem (i.e., the maximization of the acquisition function).
They achieve the extension to the parallel setting through the introduction of a frontier of promising candidates to fill in a parallel queue.

Ensemble-based asynchronous methods have each member independently assign new points to evaluate to parallel workers; therefore, the quicksand problem does not arise.
In \cite{egele2022asynchronous}, authors propose an ensemble approach deployed on multiple HPC nodes.
Each node runs its own independent BO instance with a Random Forest Regressor prior and the Upper Confidence Bound (UCB) acquisition function, each with a different value for the UCB hyperparameter.
Specifically, they perform iid sampling of the UCB trade-off parameter from an exponential distribution and assign each sampled value to one of the ensemble members.
They then communicate their results to each other asynchronously, which will be integrated into the posterior at the next possible update.

Another possible approach for asynchronous problems to escape quicksand is temporarily introducing placeholder information into the history of evaluated points so that points chosen sequentially can be derived from different posterior distributions.
This trick happens by imputing a specific value to the pending evaluation of the objective function, which the actual value will replace once the evaluation has been completed.
Two common ways to implement this approach are the Constant Liar, which always imputes a fixed constant value (computed, e.g., as the average of values obtained from previous experiments), and the Kriging Believer, which imputes the current value of the GP posterior mean $\mu_n(x)$ computed at the pending configuration \cite{ginsbourger2010kriging}.

Other approaches include \cite{kandasamy2018parallelised}, in which authors adapt Thompson sampling (TS), i.e. iteratively sampling from the posterior GP, to the parallel BO setting (both synchronous and asynchronous).
In the asynchronous case, whenever a parallel worker is available, the method computes the posterior GP given the current history of evaluated points, draws a random sample from it to compute the current acquisition function, and then maximizes it to choose the next point to allocate to a worker.
This method circumvents the quicksand issue using a randomized selection of points, unlike the usual deterministic algorithms.
Finally, \cite{snoek2012practical} suggests a sequential, fully Bayesian approach that can work with asynchronous settings.
Given a set of pending evaluations, the method integrates the base acquisition function over all possible values of these evaluations according to their current joint posterior distribution.
It then approximates this integral via Monte Carlo estimation and computes the next point to evaluate by maximizing the resulting ``integrated acquisition function.''
Since this method includes pending points in the computation of each new point, it avoids the quicksand problem.

\section{The LiGen virtual screening application}  \label{sec:ligen}

The virtual screening stage ranks the most promising ligands according to their strength of interaction with the target protein(s).
To compute this value, we first need to estimate the 3D displacement of the ligand atoms when they interact with the protein, i.e., to dock a ligand.
Then, we can consider chemical and physical interactions to summarize the interaction strength with one number, i.e., we score a ligand.
While the latter is a model we apply to evaluate a ligand-protein pair, the former is the most complex task.
Due to the difference in size between a ligand and a protein, multiple protein areas are suitable to hold a ligand, named docking sites.
Moreover, a molecule is not a rigid body, but it is possible to rotate a subset of the molecule bond, changing its shape without changing its chemical properties.
The most accurate software that docks a ligand performs a molecular dynamic simulation to let the protein-ligand system evolve until it reaches an equilibrium.
However, these computations are so expensive that their performance is expressed in nanoseconds of simulation per day \cite{kutzner2015best}.
For this reason, most docking software for virtual screening avoids the dynamics and tries to predict the final pose of the ligand using heuristics and stochastic approaches.
To improve the robustness of the solution, they usually produce multiple poses of the ligand and let the scoring function select the best one.

This paper focuses on the LiGen software as a real-world application case study.
LiGen is the core of the EXSCALATE virtual screening platform hinging on modern supercomputer nodes.
In particular, it can hinge a highly optimized CUDA implementation of the computation kernels to offload computation \cite{9817028,10.1145/3235830.3235835,vitali2022gpuoptimized} and it performs efficient I/O operations \cite{9651263}.
In a previous work \cite{gadioli2021tunable}, we show how the end-user can leverage the accuracy-performance trade-off by exposing and tuning the software knobs at runtime.
Those results, obtained using a mini-app with a simplified quality metric, showcase the possibility of maximizing the accuracy given a computation budget.
In this paper, we focus on LiGen, which exposes a more significant number of parameters and uses an asynchronous approach to offload computation.
We are also interested in a more complex definition of results quality.
Therefore, in this work, we continue on the same path, focusing on parameter exploration rather than how to expose and use them.
In the following subsection, we summarize the metrics of interest and the software knobs that we address in this paper. 

\subsection{Metrics of interest}  \label{subsec:metrics}
The virtual screening task should complete the computation in the shortest time.
The input data are available since the beginning of the execution, so the primary performance metric is the throughput in terms of the number of ligands computed in a second or, equivalently, the wall time required to virtual screen the input data.
In the current version of LiGen, we use a batched approach \cite{vitali2022gpuoptimized} that computes bundles of ligands in parallel on the GPU.
While the latter yields a greater overall throughput than spreading the computation of a single complex on the GPU, it requires more data to reach peak performance.
When considering the computation quality, its definition depends on the task.
In the application domain, the quality of molecular docking depends on the ability of the software to reproduce experimental data.
Some datasets, such as the PDBbind refined set \cite{su2018comparative}, provide a collection of complexes with a co-crystalized ligand.
The main idea is that the software shall re-dock the associated ligand and measure the Root Mean Square Deviations (RMSD) of atomic positions between the computed and observed poses.
With this quality measure, lower is better.
This paper considers the RMSD values within a representative dataset as part of our target metric of interest.
When we consider the molecular scoring task, the quality of results depends on how close we are to reproducing the experimental binding affinity.
However, we only measure the docking quality since no parameters influence this task.
Section \ref{subsec:exper-settings} details the dataset used in the experiments.

\subsection{Application structure and software knobs}  \label{sec:knobs}

\begin{figure}[t]
    \centering
    \includegraphics[width=0.75\linewidth]{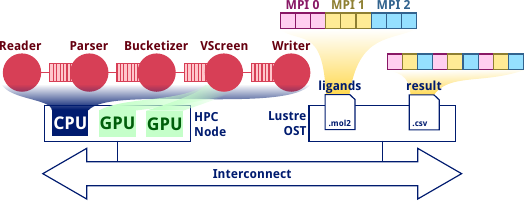}
    \caption{The LiGen application: all the MPI instances implement the same computation pipeline that works on different slabs of the input and output files. The virtual screening stage can offload computation to the node GPU(s). }
    \label{fig:ligen}
\end{figure}

\begin{table}[b]
\caption{The list of exposed software knobs and their properties.}
\label{tab:parameters}
\centering
\scriptsize
\def\arraystretch{1.2}
\begin{tabular}{lllr}
\toprule
\textbf{Knob Name} & \textbf{Affected metrics} & \textbf{Range} & \textbf{Dynamism} \\
\midrule
align\_split & Throughput and quality & [8 \dots 72] & Compile time \\
optimize\_split & Throughput and quality & [8 \dots 72] & Compile time \\
repetitions & Throughput and quality & [1 \dots 5] & Compile time \\
cuda\_threads & Throughput only & [32 \dots 256] & Runtime \\
num\_restarts & Throughput and quality & [32 \dots 256] & Runtime \\
clipping & Throughput and quality & [10 \dots 256] & Runtime \\
sim\_thresh & Throughput and quality & [1 \dots 4] & Runtime \\
buffer\_size & Throughput only & [1MB \dots 20MB] & Runtime \\
\midrule
\end{tabular}
\end{table}

LiGen implements an asynchronous pipeline, where each stage takes a ligand from an input queue, carries out a task, and then enqueues the ligand in the following stage.
The queues have a maximum size.
Thus, the slowest stage will starve the following stages and generate back pressure on the previous ones.
Each stage can use more than one system thread, scaling up to the available resources.
The virtual screening stage is usually the bottleneck, requiring all the available computation resources.

Figure \ref{fig:ligen} provides an overview of LiGen, which is an MPI application.
We use MPI to replicate the same pipeline on all the nodes that contribute to the computation.
Since virtual screening is an embarrassingly parallel problem, we must synchronize only the I/O toward and from the storage.
Figure \ref{fig:ligen} reports an example with three MPI processes represented by different colors.
We use a static partition of the input file according to its size and the overall number of MPI processes.
Each process issues independent I/O requests to read a chunk of data, so the reading pace depends on the slowest stage of the pipeline.
On the other hand, we need to synchronize the write operation to prevent one process from overwriting the output of another one.
Moreover, it is possible to aggregate the output in fewer processes, named writers, to reduce the pressure on the file system.
Besides synchronizing the collection and aggregation of output data, each writer issues independent I/O requests to the file system.
The frequency of the write operations depends on the slowest MPI process.
A previous work \cite{9651263} describes more technical details about the LiGen IO.

The virtual screening pipeline is composed of the following stages.
The reader stage performs the I/O requests to read the ligands from the chemical library, encoded in a textual representation.
The parser stage populates the internal data structure using the textual representation.
During a bucketizer stage, ligands are bundled according to input features such as the number of atoms and rotamers.
This out-of-order execution improves the GPU computation \cite{vitali2022gpuoptimized}.
The next stage carries out the virtual screening computation.
It is possible to choose the implementation we want to use at runtime, and in this paper, we focus on the CUDA implementation.
The writer stage collects the results in an accumulation buffer and issues I/O requests to store them in the file system.
In the ideal scenario, the virtual screening stage is the bottleneck, which means that all the computation resources of a node are busy carrying out useful computation.

When focusing on the virtual screening stage, we use a gradient descent algorithm with multiple restarts.
At each restart, we generate a different ligand pose. 
In the first phase, we align the pose to the pocket using rigid movement.
Then, we optimize its shape to match the target pocket.
We can score either all generated poses or just a subset to reduce the computational effort.
The final ligand score is the value of the best pose that we have found.

Table \ref{tab:parameters} lists all the exposed software knobs that alter the application execution.
In particular, the \texttt{align\_split}, \texttt{optimize\_split}, and \texttt{repetitions} parameters define how thoroughly we perform the gradient descent.
The \texttt{num\_restarts} parameter states how many poses we generate for each ligand, while the \texttt{clipping} parameter states how many poses we score.
Finally, the \texttt{sim\_thresh} parameter defines a threshold that LiGen uses to deem two ligand poses as different.
We use this concept to cluster poses together when we select which pose we need to score to promote heterogeneity \cite{gadioli2021tunable}.
Some of them are runtime parameters, while others are compile-time knobs.
When we need to evaluate several LiGen configurations with different values for compile-time knobs, we have to compile the application from scratch.
All these parameters affect the application throughput and quality.
Larger values lead to a more fine-grained computation, increasing the result quality.

A second set of parameters defines the application pipeline and aims to tailor its execution to the underlying hardware.
For this reason, they only affect the application throughput.
In particular, \texttt{cuda\_threads} defines how many threads we use in a CUDA block, while the \texttt{buffer\_size} parameter selects the size of each write operation toward the file system.

We also want to guarantee a minimum quality of the result regarding RMSD.
For this reason, we place a constraint on the measured RMSD on the representative dataset, requiring that it be smaller than $R_{max}$.
We define LiGen configurations that do not meet this criterion as unfeasible.
We will provide more details about the constraint formulation and unfeasible configurations in Section \ref{subsec:autotuning-ligen-parallel-bo}.

\section{Application Tuning}  \label{sec:autotuning}
This section describes the main components of our autotuning methodology.
Section \ref{subsec:framework-architecture} overviews the architecture of our framework to execute LiGen and collect information for the subsequent configurations to evaluate in an HPC setting.
In Section \ref{subsec:autotuning-ligen-eval}, we describe how we evaluate the extra-functional properties of a LiGen configuration.
Section \ref{subsec:autotuning-ligen-parallel-bo} explains the parallel BO algorithms we use to conduct black-box optimization of LiGen.

\subsection{Framework architecture}  \label{subsec:framework-architecture}
\begin{figure}[t]
    \centering
    \begin{subfigure}[b]{0.45\linewidth}
        \centering
        \includegraphics[width=0.9\linewidth]{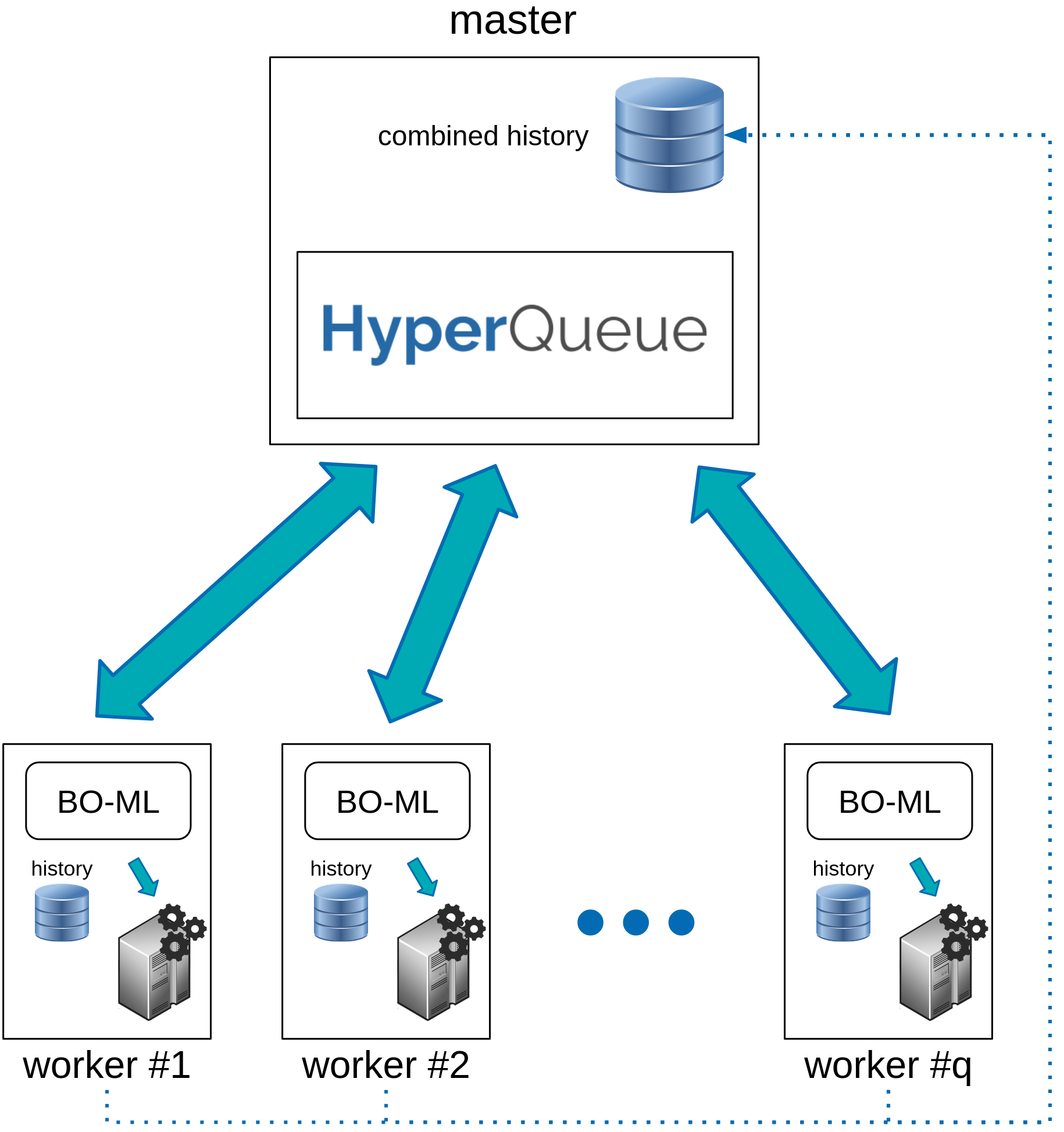}
        \caption{Distributed architecture.}
        \label{subfig:architecture-emaliboo}
    \end{subfigure}%
    \begin{subfigure}[b]{0.45\linewidth}
        \centering
        \includegraphics[width=0.9\linewidth]{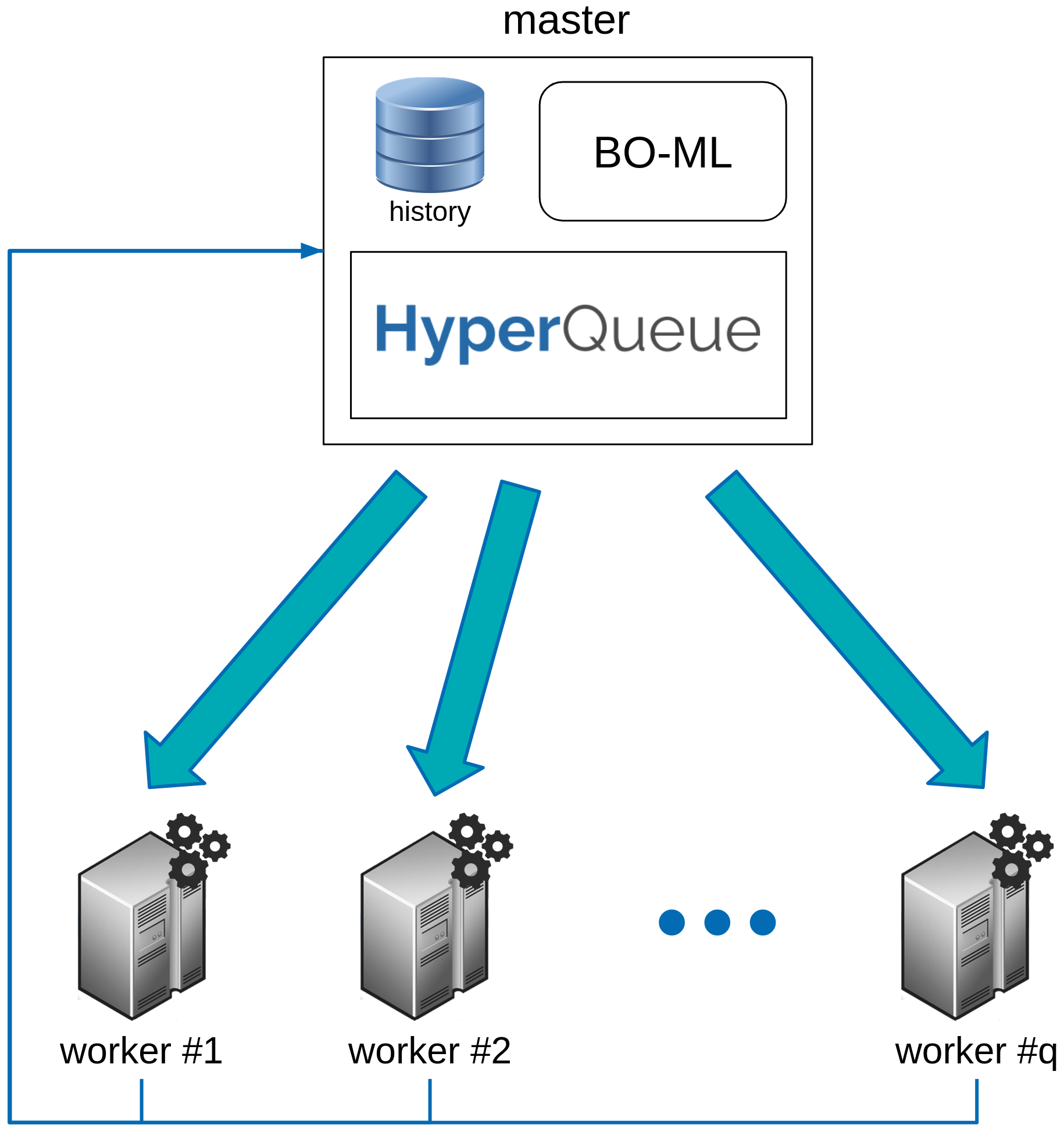}
        \caption{Centralized architecture.}
        \label{subfig:architecture-pamaliboo}
    \end{subfigure}
    \caption{Framework architecture.}
    \label{fig:architecture}
\end{figure}
We display the architecture of our autotuning framework to execute and optimize LiGen in Figure \ref{fig:architecture}.
One can deploy the framework in two different settings: with a distributed architecture (see Figure \ref{subfig:architecture-emaliboo}) and with a centralized one (see Figure \ref{subfig:architecture-pamaliboo}).
Each of these settings fits one of the two variations of parallel asynchronous BO presented in Section \ref{subsec:autotuning-ligen-parallel-bo}, i.e., EMaliboo and PAMaliboo, respectively.

Both variants exploit the HyperQueue scheduler to execute new LiGen configurations.
HyperQueue \cite{Beranek2024HQ} is a lightweight, self-contained task execution runtime designed to simplify the orchestration of large workflows on HPC clusters, managing to scale to hundreds of nodes/workers.
It initializes a \textit{server} which manages several \textit{workers} (corresponding to different cores or compute nodes), automatically asking for computational resources and performing dynamical load balancing of tasks across all allocated nodes.
It has a seamless interface with SLURM and PBS schedulers if they are present on the cluster, but it can also work without them as a general task executor.
After obtaining a new LiGen configuration to test, our framework submits the corresponding LiGen configuration to the HyperQueue server, which then sends it to a specific worker to be executed.
In particular, each HPC compute node represents a parallel worker and can execute one LiGen instance at a time.

In the distributed architecture of Figure \ref{subfig:architecture-emaliboo}, each compute node has its own separate BO engine, worker, and history and interacts with HyperQueue as an intermediary to make the worker execute LiGen with a specific configuration.
At the end of the optimization process, the framework collects the evaluated configurations from each node into a combined history of evaluations.
On the other hand, the centralized architecture of Figure \ref{subfig:architecture-pamaliboo} has a single master node that selects new LiGen configurations to test, interacts with HyperQueue, and stores a shared history of visited configurations.
All information from completed LiGen executions is communicated to the master node and contributes to selecting the following configurations.

We will delve into the algorithmic details of configuration selection via BO in Section \ref{subsec:autotuning-ligen-parallel-bo}.

\subsection{Evaluating the LiGen performance}
\label{subsec:autotuning-ligen-eval}
The LiGen configuration evaluation is not a trivial process for two reasons.
A subset of the configuration parameters are constants known at compile time that require rebuilding the executable.
Moreover, we need two completely different experiments to measure the average throughput and the quality of results.
For the throughput, we need to virtual screen a dataset with many ligands to reach the peak performance \cite{vitali2022gpuoptimized}.
As a rule of thumb, we need at least one million molecules for each available GPU.
In this case, a single LiGen run is enough and lasts a considerable amount of time.
Instead, to measure the quality, we need to perform multiple LiGen executions that re-dock a single ligand against the related binding site for all the complexes we want to include in the evaluation.
The main point is that we cannot use the throughput of the experiment to measure the quality since, for virtual screening, we avoid copying out the displacement of the atoms after the docking from the GPUs for performance reasons.
This limit denies the possibility of computing the RMSD with the ground truth.
On the other hand, when we evaluate the molecular docking quality, we compute a single molecule, making the GPU usage very inefficient.
For this reason, to evaluate a LiGen configuration we need to build the application from sources and perform two distinct experiments to measure the two extra functional properties.

Most of the parameters reported in Table \ref{tab:parameters} aim at tailoring the LiGen execution on the target HPC node without altering the molecular docking algorithm as described in Section \ref{sec:knobs}.
We can define the LiGen configuration vector $x$ as the set of parameters that only affect the throughput $x_{t}$ plus the ones that also affect the quality $x_{tq}$, i.e., $x = x_{t} \cup x_{tq}$.
With this formulation, we can write the LiGen execution time under configuration $x$ as $T(x) = T(x_t, x_{tq})$.
Similarly, we measure the RMSD between the experimental value and the closest pose LiGen generates for each element in the dataset.
Then, we compute an aggregate metric that we indicate with $R(x) = R(x_{tq})$ to represent the quality of the configuration.
In this paper, we define $R(x)$ as the 75th percentile of the RMSD values on the dataset.
It is important to note that $R(x)$ is agnostic from the underlying hardware but only depends on the molecular docking algorithm.
To take both efficiency and solution quality into account, we formulate the following optimization objective: $f(x) = R^3(x) \, T(x)$, where lower is better.
This metric allows simultaneous minimization of both the RMSD and the execution time while giving more weight to the RMSD.
We will show the proper mathematical formulation of our optimization problem in the next section.

To automatize the configuration evaluation, we use a wrapper that compiles LiGen in a dedicated directory and runs it with a specific configuration.
Finally, it measures and collects the execution time and solution quality, providing the information to the BO engine (see Sections \ref{subsec:framework-architecture} and \ref{subsec:autotuning-ligen-parallel-bo}) that orchestrates the execution.
To minimize the evaluation time in the autotuning phase, the wrapper uses a cache system to keep track of the known values of $R(x) = R(x_{tq})$.
Indeed, if a new configuration $x$ shares its $x_{tq}$ part with an already visited configuration, we need not repeat the RMSD evaluation on the entire dataset.
Moreover, when we deploy LiGen to a new HPC system, the autotuning phase can spare most of the $R(x)$ evaluations.
We publicly released the source code for this wrapper\footnote{\url{https://gitlab.com/margot_project/ligen_exploration_wrapper}}.
Even if this wrapper is tailored for LiGen, its structure can easily extend to other applications.

\subsection{Efficient multi-node configuration exploration}  \label{subsec:autotuning-ligen-parallel-bo}
To optimize LiGen via Bayesian Optimization (BO), we write a specific formulation of Equation \eqref{eq:problem-general}.
Specifically, given a LiGen configuration vector $x$, our goal is to solve the following constrained optimization problem:
\begin{align}
    \begin{split}
    \min_{x \in A} f(x) &= R^3(x) \, T(x) \\
    \text{s.t.} \quad g(x) &= R(x) \le R_{max}
    \end{split}
    \label{eq:problem-ligen}
\end{align}
where $R(x)$ and $T(x)$ are the RMSD 75th percentile and the LiGen execution time, as described in the previous section.
We also set a quality threshold for the solution associated with configuration $x$.

This work proposes two parallel extensions to the MALIBOO approach presented in \cite{guindani2024integrating} to a parallel HPC setting.
MALIBOO (MAchine Learning In Bayesian OptimizatiOn) integrates sequential BO with ML techniques to minimize the execution costs of recurring resource-constrained computing jobs.
In particular, it incorporates an ML performance model estimating the value of a constrained resource into the acquisition function (specifically, \cite{guindani2024integrating} uses Expected Improvement or EI).
MALIBOO trains this model online, using all data collected by the BO algorithm.

ML methods can be helpful in optimization problems because of their predicting capabilities.
This statement is especially true in a setting where information on the objective $f(\cdot)$ and the constraint $g(\cdot)$ is scarce, given that both are black-box, expensive-to-evaluate functions.
For instance, it is possible to convey information about the violation of constraints by guiding the search towards points $x \in A$ that most likely satisfy them.
The BO algorithm benefits from this approach, as our goal is ultimately to find optimal configurations that are also feasible, i.e., points $x$ such that $R(x) \le R_{max}$.

We now describe the two proposed parallel extensions, which have different advantages in different situations.
Both serve the purpose of finding the optimal configuration to execute LiGen in a production environment, i.e., the solution of Equation \ref{eq:problem-ligen}.
Section \ref{subsubsec:emaliboo-approach} presents EMaliboo, an ensemble-based technique best suited for expansive exploration.
An ensemble of individual agents separately exploring grants this technique a better coverage of the configuration space and, ultimately, a higher quality of proposed configurations.
This technique excels in scenarios in which smaller GP and ML models are enough to accurately estimate the objective and constraint functions $f(\cdot)$ and $g(\cdot)$.
In Section \ref{subsubsec:pamaliboo-approach}, we present PAMaliboo, a robust, centralized approach that is most useful in scenarios where large amounts of data are needed to perform predictions.
Since this technique collects all currently available information to build a single model, it is more appropriate when the models require larger amounts of data to perform meaningful estimation.

\subsubsection{Ensemble approach: EMaliboo} \label{subsubsec:emaliboo-approach}
The first proposed method is EMaliboo, an Ensemble-based extension of the MALIBOO sequential technique.
We deploy it with the distributed architecture of Figure \ref{subfig:architecture-emaliboo}.
Assuming that one master node and $q$ other nodes (i.e., the parallel workers) are available, this technique exploits an ensemble of $q$ independent agents, each running their instance of sequential MALIBOO.
Besides the starting random seed for the choice of initial points, the agents behave identically.
In particular, they all use the ML-integrated acquisition function described in \cite{guindani2024integrating}.
This acquisition function contains an ML model $\hat{g}(x)$ estimating the black-box constraint function, which in our case is $R(x)$.
These agents explore the LiGen search space on their own, choosing new configurations and executing the application with them, but they do not share information.
We choose to keep the agents' histories separate, i.e., make the agents independent, to grant the ensemble greater exploration.
In this way, the exploration of a particular area by one of the agents does not influence another agent.
This independence allows each parallel worker to remain busy at all times (besides the short overhead to compute new configurations to test), exploiting the maximum possible amount of computational resources on the HPC system.

At the end of the optimization procedure, we collect all points visited by the ensemble agents.
The best configuration found by the algorithm is the point associated with the best value of the objective function found by any of the agents, i.e., the best point among the candidates of the individual agents.

\subsubsection{Pure Asynchronous approach: PAMaliboo} \label{subsubsec:pamaliboo-approach}
The second extension is PAMaliboo, a Pure Asynchronous parallel extension to MALIBOO.
PAMaliboo also leverages an ML-integrated acquisition function $a(x)$ to solve the optimization problem in Equation \eqref{eq:problem-ligen}.
We run this algorithm with the deployment depicted in Figure \ref{subfig:architecture-pamaliboo}.
In particular, a single, centralized agent chooses sequentially new configurations to evaluate by maximizing $a(\cdot)$ once for every iteration and immediately submits (via HyperQueue) each new configuration to one parallel worker for evaluation in an asynchronous fashion.
To get around the quicksand issue described in Section \ref{subsec:backgr-bo-algos}, we adopt a Kriging Believer approach as suggested by \cite{ginsbourger2010kriging}.
That is, after the master sends a new configuration $x_{n+1}$ to a parallel worker for evaluation, we add a temporary placeholder point to the history of evaluated points: $(x_{n+1}, \mu_n(x_{n+1}))$, i.e., we set the objective function evaluation equal to the current posterior mean at that configuration.
This action only happens when there is at least one idle parallel worker.
When one or more evaluations have been completed, we remove all placeholder points from the history and add the correct evaluations.
Once every $P$ iterations (where $P$ is a hyperparameter), we also re-train the ML model $\hat{g}(x)$, estimating the constraint.

We summarize this procedure in the pseudo-code of Algorithm \ref{algo:pamaliboo}.
\begin{algorithm}[htb!]
    \caption{PAMaliboo algorithm}
    \label{algo:pamaliboo}
    \SetAlgoVlined
    choose $N$ iterations number, $polling$ time, $P$ training periodicity\;
    choose $n_0$ initial points\;
    evaluate $f(\cdot)$ in the initial points synchronously, add all evaluations to history $H$\;
    fit GP model with data in $H$\;
    initialize current iteration $n=0$\;
    \While{$n < N$}{
        \If{there are completed evaluations}  {  
            remove all placeholder values in $H$\;
            add all completed evaluations to $H$\;
        }  \label{algo:pam:compl-eval-endif}
        \eIf{maximum parallelism reached}  {  
            sleep for $polling$ seconds\;  
        } {
            \If{$n\% P == 0$}  {  
                train ML model in $a(\cdot)$ for the constraint with all true constraint data\;  
            }
            find point $x_{n+1}$ maximizer of $a(\cdot)$ under the current model\;  
            submit evaluation of $f(x_{n+1})$ asynchronously\;
            add placeholder value $(x_{n+1}, \mu_n(x_{n+1}))$ to $H$\;  
            update the current posterior distribution of the GP model with data in $H$\;
            $n++$\;  
        }
   }
   return estimated optimum $\hat{x} = \argmin_{x \in H} f(x)$
\end{algorithm}
The algorithm is similar to sequential BO.
At each iteration, before proceeding with the exploration, we first gather completed evaluations of the objective functions (if any) and remove all placeholder values from history $H$ (steps 7 to 10).
We then wait for $polling$ seconds before starting over if all parallel workers are busy (steps 11--12).
In such a way, we never create a waiting queue for candidate configurations.
Indeed, we would instead submit a configuration using all available information at the time of submission instead of queueing a configuration whose choice may depend on information potentially outdated shortly.
The algorithm then proceeds as regular BO, with the difference that we add a placeholder value to the history (step 19) rather than the actual function value since the latter is currently unavailable.
Also, we update our ML model on constraints once every $P$ algorithm iterations (step 15).

\begin{figure}[t]
    \centering
    \begin{subfigure}[b]{0.5\linewidth}
        \centering
        \includegraphics[width=\linewidth]{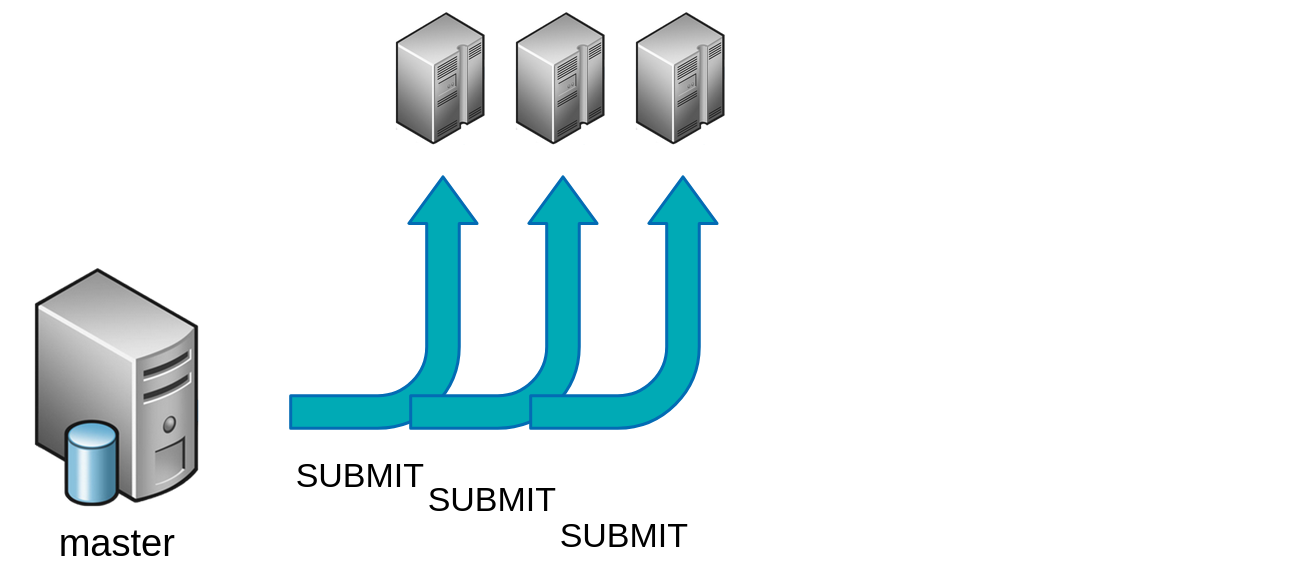}
        \caption{Three jobs are submitted back to back.}
        \label{subfig:pamaliboo-1}
    \end{subfigure}%
    \begin{subfigure}[b]{0.5\linewidth}
        \centering
        \includegraphics[width=\linewidth]{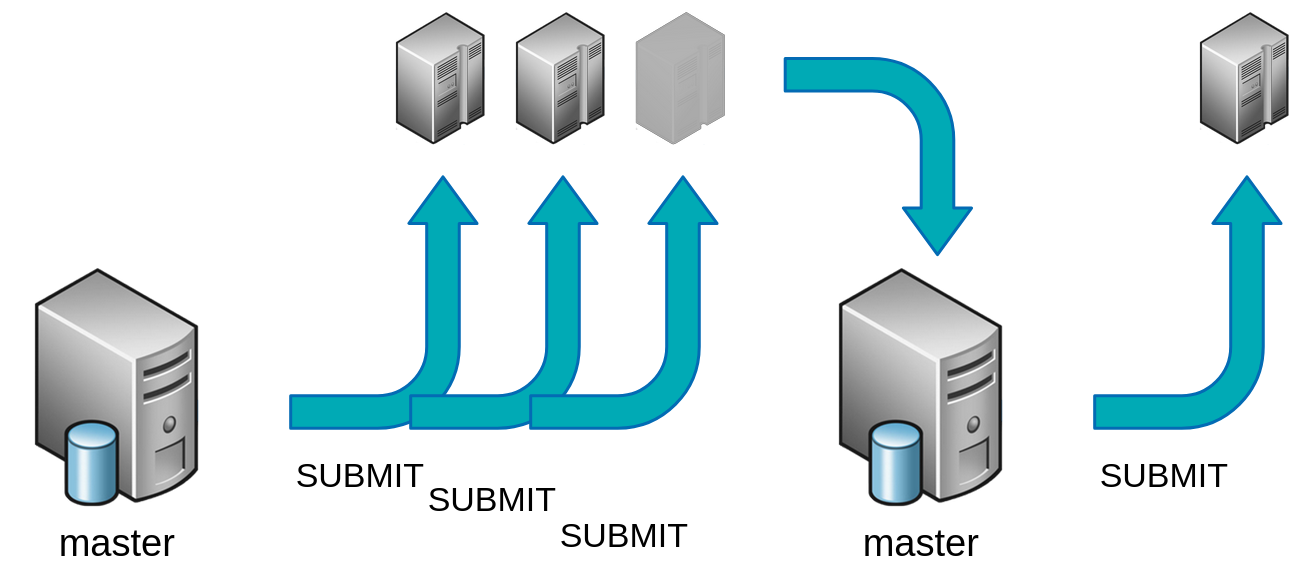}
        \caption{Immediately after one of the three jobs finishes, a new one is submitted.}
        \label{subfig:pamaliboo-2}
    \end{subfigure}
    \caption{PAMaliboo technique with $q=3$ parallel workers.}
    \label{fig:pamaliboo}
\end{figure}

We show an example behavior of the system in Figure \ref{fig:pamaliboo}, with $q=3$ parallel workers.
Thanks to the placeholder value mechanism, BO can obtain new configurations to test immediately without waiting for previous jobs to complete.
It submits new jobs one after another, as long as there are idle parallel workers, as seen in Figure \ref{subfig:pamaliboo-1}.
As soon as one of the previous jobs finishes (with a delay of $polling$ seconds at most, see Algorithm \ref{algo:pamaliboo}) and a worker becomes idle, the framework recovers the execution results, updates the models, and then selects a new configuration to submit a job with, as represented in Figure \ref{subfig:pamaliboo-2}.
In this way, we can exploit the total computational power of the system as much as possible, similarly to the other algorithm described earlier.

Unlike the ensemble-based EMaliboo, this method exploits a single centralized ML model for constraint prediction and a single Gaussian Process (GP).
Therefore, it better capitalizes on the acquired constraint information than the independent agents, each of which individually acquires less data and might, thus, be less accurate in predicting the constraint value.

\section{Experimental results}  \label{sec:exper}
We now describe the context and results of our experimental campaign, in which we run the two proposed BO-based algorithms and one state-of-the-art competitor.
These techniques execute the LiGen application multiple times on parallel nodes, using the deployments described in Section \ref{subsec:framework-architecture} and Figure \ref{fig:architecture}.
Some experiments use a simulated version of LiGen in a controlled environment, while others execute the actual application in a prototype environment.
All algorithms aim to find the best feasible configuration for LiGen, where ``best'' means with the smallest value of the objective function $f(\cdot)$ and ``feasible'' means that it must respect the RMSD constraint, as formalized in Equation \ref{eq:problem-ligen}.
For this reason, we measure the efficiency of an algorithm with the simple regret, namely the distance in terms of $f(\cdot)$ values between the best solution found by the algorithm and the actual best configuration.
On the other hand, we will evaluate the accuracy of ML models trained by our proposed algorithms via their Mean Absolute Percentage Error (MAPE) computed on newly selected points.
We also conduct simulated experiments in which we inject an error into the aforementioned ML models to observe the change in the behavior of our algorithms in the presence of significant prediction errors.

We describe our competitor approach from the literature, OpenTuner, in Section \ref{subsec:exper-comp-ref}.
Then, we cover the algorithmic settings of our experiments in Section \ref{subsec:exper-settings}.
In the subsequent sections, we describe the settings and results of several categories of experiments: simulated experiments in Section \ref{subsec:exper-simul}, analyses with error injection in Section \ref{subsec:exper-err-inj}, and experiments in the prototype environment in Section \ref{subsec:exper-real}.
Finally, we summarize and discuss the results in Section \ref{subsubsec:exper-discussion}.

\subsection{Comparison reference approach}  \label{subsec:exper-comp-ref}
We compare our techniques against OpenTuner \cite{ansel2014opentuner}, a popular open-source ensemble-based autotuning meta-technique.
It uses an ensemble of multiple search techniques to conduct black-box constrained or unconstrained optimization of the given application.
A meta-search algorithm guides the exploration process, allocating more tests to techniques that perform well.
The techniques coordinated by the meta-algorithm include a suite of classical optimization methods such as differential evolution and greedy mutation variants.
Individual techniques share results through a common database so that improvements made by one of them can also benefit the others.
This sharing occurs in technique-specific ways; for example, evolutionary techniques add results found by other techniques as members of their population.
The default meta-algorithm in OpenTuner is the ``multi-armed bandit with sliding window, area under the curve credit assignment,'' also known as AUC Bandit, coordinating the DifferentialEvolutionAlt, UniformGreedyMutation, NormalGreedyMutation, and RandomNelderMead search techniques.
The ensemble of several methods allows extensive exploration of the optimization domain, while the clever allocation of tests by the meta-technique encourages the exploitation of promising paths.

\subsection{Dataset and algorithm settings}  \label{subsec:exper-settings}
The LiGen performance, both in terms of quality and throughput, depends on features of the input \cite{9817028}.
If we focus on the throughput, we selected 200 molecules from the MEDIATE dataset\footnote{\url{https://mediate.exscalate4cov.eu}} with an even distribution of heavy atoms and rotamers.
To measure a steady throughput without the initial transitory phase \cite{vitali2022gpuoptimized}, we replicate these molecules 10000 times.
The dataset we use to measure the throughput has 2 million molecules.
Since the throughput does not depend on the features of the binding site, we provide a single target.
When measuring the quality of results, we use the refined set of the PDBbind \cite{su2018comparative}.
It contains the atom displacement of four thousand complexes that we can use to compute the error distribution in terms of RMSD.

We now discuss the settings for all algorithms involved in the experimental validation.
All algorithms exploit a number of parallel workers $q$, also known as \textit{parallelism level}, equal to 10.
Each algorithm receives $n_0 = 30$ randomly selected initial points and runs for $N = 1000$ iterations (i.e., it conducts 1000 evaluations of the objective function, each representing an individual execution of LiGen).
For the ensemble-based EMaliboo, we partition these points evenly on its parallel independent workers, i.e., each has 3 initial points and runs for 100 iterations.
For both techniques, we set the training periodicity $P$ of our ML models to 3 and the polling interval to 1 second (see Section \ref{subsubsec:pamaliboo-approach}).
Finally, for all algorithms, we run the same experiments with 5 or 10 different random seeds (depending on the type of experiment) to ensure the statistical robustness of our results.
All algorithms receive the same set of initial points for each seed.

We set the RMSD threshold in Equation \ref{eq:problem-ligen} to $R_{max} = 2.1$, which is a significant value in the distribution of RMSD values according to preliminary analyses.
Other settings, such as Expected Improvement as the acquisition function of choice and a Ridge regression model as the constraint estimator $\hat{g}(\cdot)$, are similar to the original MALIBOO technique \cite{guindani2024integrating}.
As a linear regression-based model, Ridge behaves fundamentally differently from the decision tree-based XGBoost, the model generating the simulated LiGen data, allowing for a fair prediction, as we will discuss in the next section.

\subsection{Simulated experiments}  \label{subsec:exper-simul}
If we want to perform black-box optimization of the LiGen application, one has to deal with its significant execution times, which span up to 40 minutes for a single execution.
Running multiple optimization processes that execute such applications hundreds or thousands of times would be unfeasible compared to our time, budget, and computational power constraints.
For these reasons, we conduct preliminary experiments with a simulated version of LiGen, represented by a pair of oracle-like ML models.
Given a LiGen input configuration $x$, these models output the corresponding execution time $T(x)$ and the RMSD percentile $R(x)$, respectively (see Equation \ref{eq:problem-ligen}).
For training these ML models, we use a dataset of about 1000 entries by collecting all configurations previously tested on the actual LiGen application, along with their corresponding execution times $T(x)$ and RMSD values $R(x)$.
We conducted this profiling on the prototype environment described in Section \ref{subsec:exper-real} and lasted about two weeks.
The collected dataset represents our previous knowledge of LiGen, consisting of actual metrics evaluated by running the real software in a test environment.
We then performed independent offline training of separate ML models for predicting $T(\cdot)$ and $R(\cdot)$ values via aMLLibrary \cite{guindani2023amllibrary} by using the aforementioned dataset and performing an 80-20 train-test split.
The trained models were Random Forest, Support Vector Regression (SVR), and XGBoost.
The best-performing prediction model for both metrics was XGBoost, scoring low test-set MAPEs of less than 5\%.
These results show that the two XGBoost models are an accurate replacement to simulate the execution of LiGen for our optimization experiments.

We run our simulated experiments on a private server with Ubuntu 22.04, 1 TB RAM, and 40 Intel(R) Xeon(R) Silver 4316 CPU @ 2.30 GHz with 2 threads per core.
The processing time of our BO-based algorithms, i.e., the overhead for choosing a new configuration, is between 15 and 25 seconds, increasing with the iterations number (as the algorithm visits more and more points to include in the computation).
The bulk of this overhead comes from the maximization of the acquisition function, for which we use Nelder-Mead heuristics with multiple restarts \cite{nelder1965simplex}, similarly to \cite{guindani2024integrating}.
Nonetheless, these time lapses are minor compared to the LiGen execution times.
On the other hand, OpenTuner has an algorithm overhead close to zero due to the simple computations required by the meta-technique to allocate a new test.

\begin{figure}[t]
    \centering
    \newcommand\mult{0.45}
    \begin{subfigure}[b]{\mult\linewidth}
        \centering
        \includegraphics[width=\linewidth]{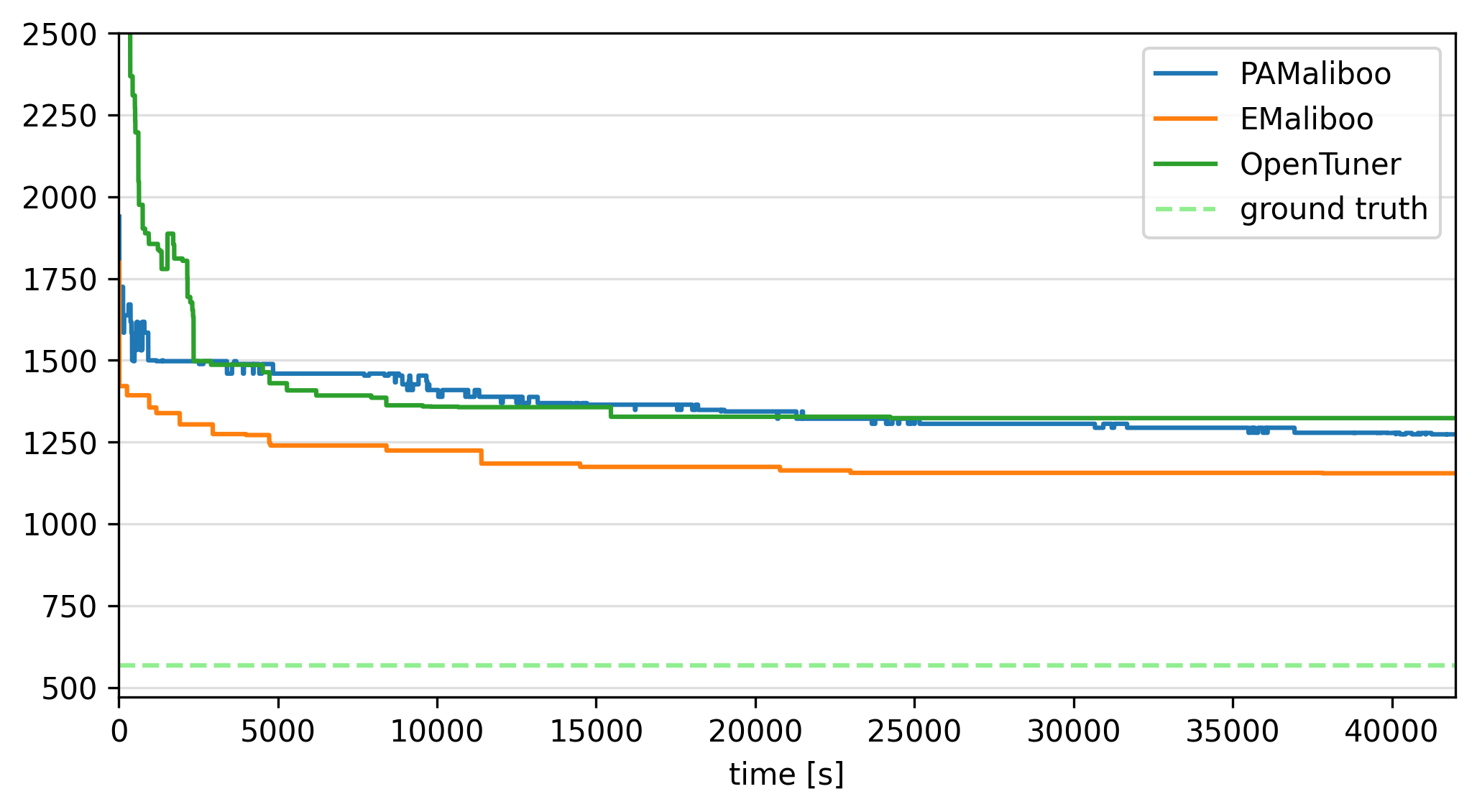}
        \caption{Current best feasible value of the objective function over time.}
        \label{subfig:results-base-a}
    \end{subfigure}
    \begin{subfigure}[b]{\mult\linewidth}
        \centering
        \includegraphics[width=\linewidth]{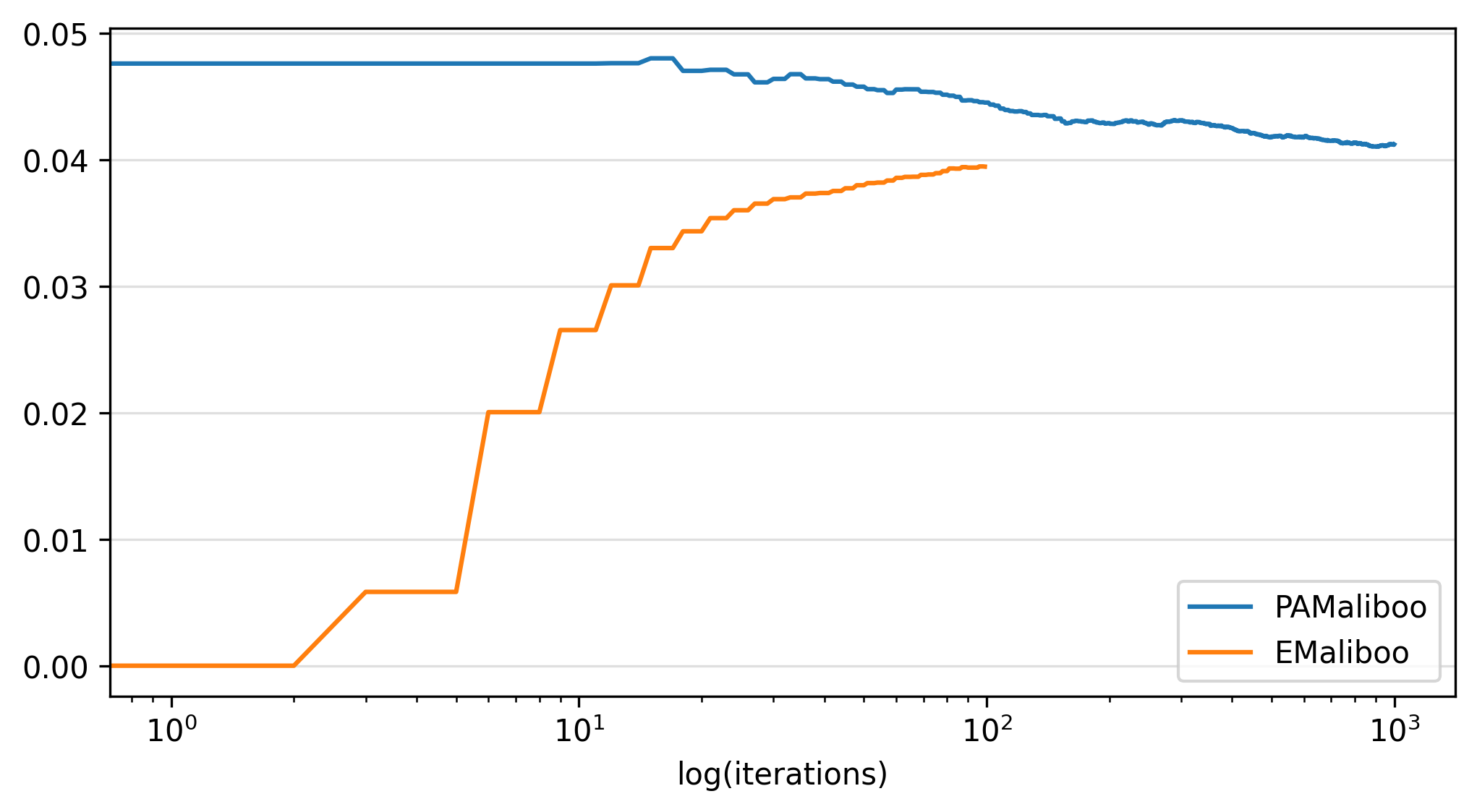}
        \caption{MAPE of ML models over log-scale of iterations.}
        \label{subfig:results-base-b}
    \end{subfigure}
    \begin{subfigure}[b]{\mult\linewidth}
        \centering
        \includegraphics[width=\linewidth]{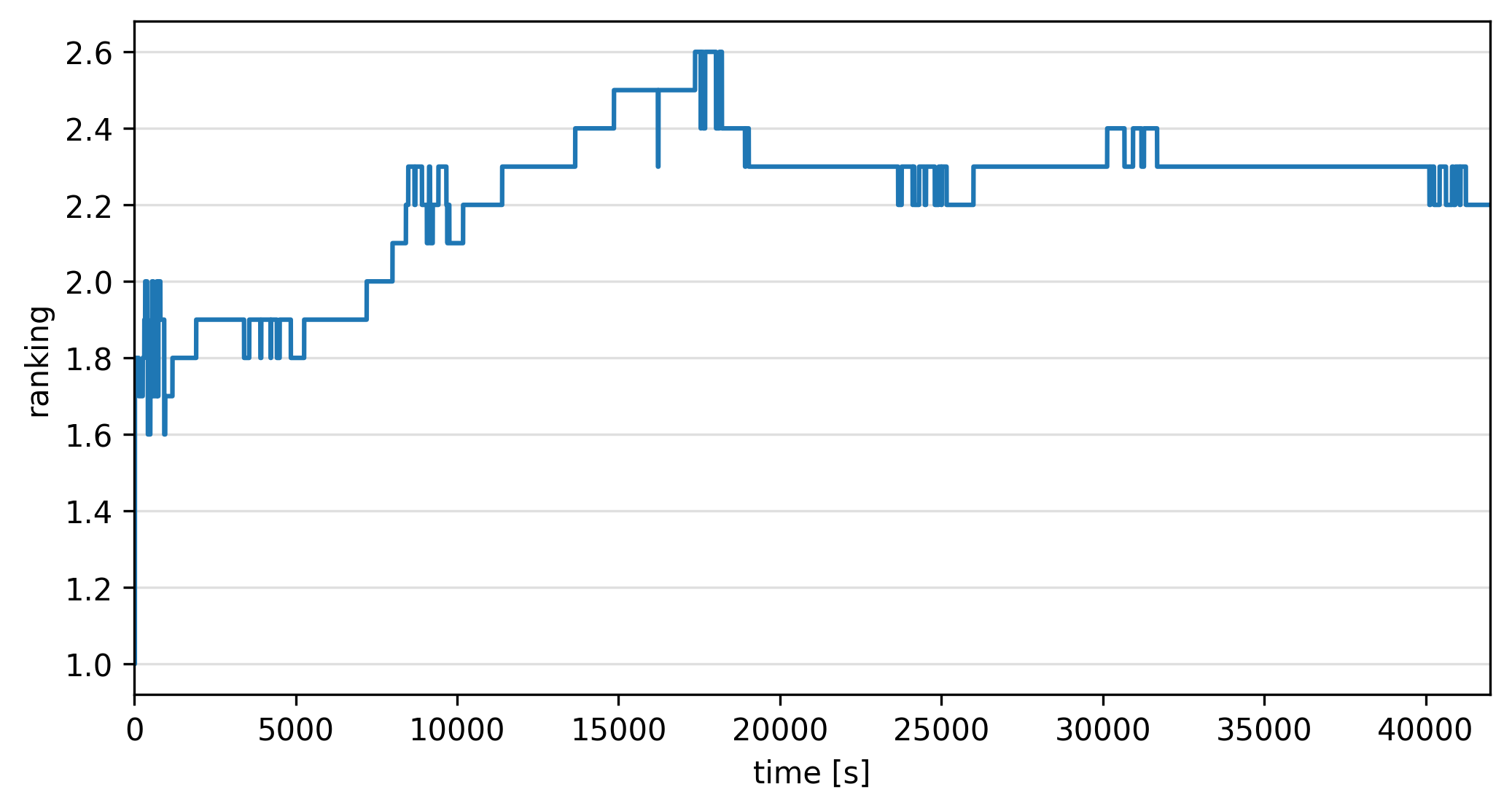}
        \caption{Ranking of PAMaliboo agent vs EMaliboo members over time.}
        \label{subfig:results-base-c}
    \end{subfigure}
    \caption{Simulated experiments: 10-experiment average of relevant quantities.}
    \label{fig:results-base}
\end{figure}

We show three representative plots summarizing our experiments in Figure \ref{fig:results-base}.
We represent metrics averaged over the 10 differently seeded runs in each plot.
In Figure \ref{subfig:results-base-a}, we represent the evolution of the \textit{feasible regret} over time.
For each instant in time, we represent the best (smallest) value of the objective function $f(x)$ found up until that time corresponding to a configuration that satisfies the RMSD constraint and measure its difference with the ground-truth optimal solution.
In this case, the optimum was found by inspection, evaluating all possible configurations with the oracle-like XGBoost models and finding the one associated with the smallest simulated $f(x)$ value.
The blue line represents the centralized PAMaliboo, the orange line represents the ensemble-based EMaliboo, and the green line represents the OpenTuner framework.
Note that at any point throughout the optimization procedure, the regret of the ensemble is equal to the best (smallest) regret found by any of the ensemble agents.
In Figure \ref{subfig:results-base-a}, we also show the ground-truth optimum of our synthesized version of LiGen with a dashed green horizontal line.

This plot shows that over multiple algorithm runs, the ensemble-based EMaliboo technique consistently outperforms centralized PAMaliboo and state-of-the-art OpenTuner in terms of feasible regret, respectively, by 10\% and 13\% at the end of the observed time window.
EMaliboo converges to a better configuration, i.e., one with a smaller value of the objective function, and also has a shorter transitory phase than the other algorithms, as evidenced by the steepness of the initial part of the regret curve in Figure \ref{subfig:results-base-a}.
This technique performs better than PAMaliboo due to its extensive exploration capabilities, which are granted by the presence of multiple individual agents.
Indeed, previous experience with LiGen \cite{guindani2024integrating} shows that the large size of its domain rewards exploration-heavy tuning techniques.

In Figure \ref{subfig:results-base-b}, we plot the MAPE of the model $\hat{g}(x)$ estimating the constrained resource, computed on the newly chosen configuration $x$, over the course of the algorithm iterations (note that the abscissa is in log-scale).
As mentioned, EMaliboo agents only run for 100 iterations each, and the represented MAPE at iteration $i$ is the average of the individual MAPE values of ensemble members at their $i$th evaluated point.
We do not show OpenTuner since it uses no ML performance model. 
Although EMaliboo runs for fewer iterations, both BO-based techniques stabilize at around 4\% prediction error, indicating that their ML models have strong prediction capabilities.
In particular, the EMaliboo agents start with very few initial points, allowing their ML models to have an extremely low MAPE since these points have low variance overall.
As the algorithm visits more and more diverse configurations, the error slightly increases until it stabilizes at a more realistic threshold.
Figure \ref{subfig:results-base-b} shows that a large, centralized model such as PAMaliboo is not needed to perform accurate predictions of the constrained resource and that the smaller models with fewer points in EMaliboo are enough to estimate the feasibility of future configurations reliably.
The same holds for the GP modeling the objective function.
This observation is consistent with the good performance of EMaliboo in terms of feasible regret.

Finally, in Figure \ref{subfig:results-base-c}, we show the ranking of the PAMaliboo centralized agents compared to the individual EMaliboo ensemble members regarding feasible regret.
In this context, at a given time instant, ``ranking'' means the relative position of the PAMaliboo agent compared to the 10 ensemble members according to their feasible regret values at that instant.
Similarly to the rest of Figure \ref{fig:results-base}, we plot the average ranking across 10 different algorithm runs, which explains why fractional values appear.
The subfigure illustrates that the single-agent PAMaliboo, although inferior to EMaliboo regarding feasible regret, is still competitive with the latter, often scoring second place (out of 11) compared to the ensemble members.

\begin{figure}[htbp]
    \centering
    \includegraphics[width=0.65\linewidth]{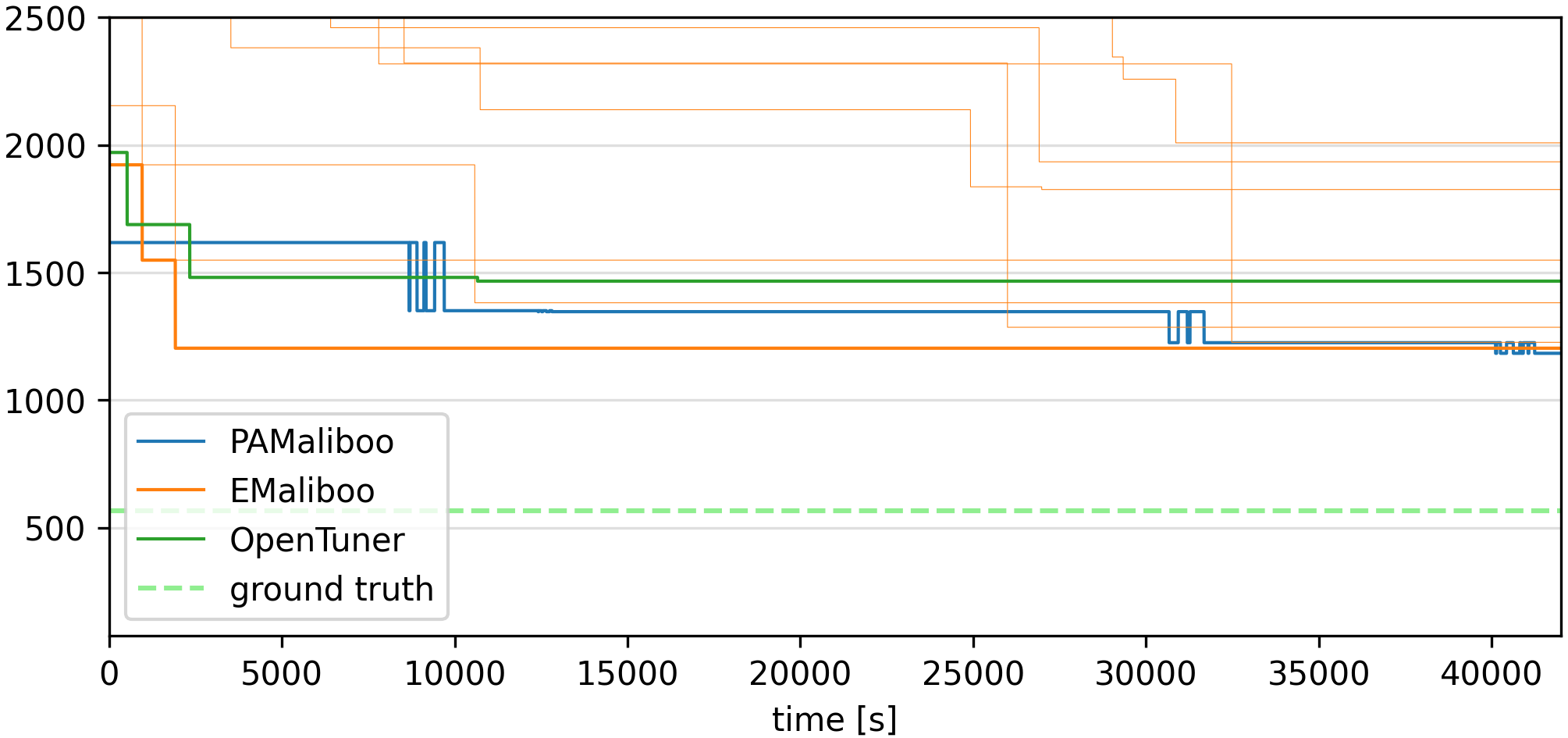}
    \caption{Simulated experiments: regret plot of one representative experiment.}
    \label{fig:results-base-individual}
\end{figure}

We have shown the average metrics over 10 differently seeded experiments.
We now focus on one individual representative experiment among those 10 to showcase how the ensemble-based EMaliboo performs exploration.
In particular, we show the simple-regret plot of this experiment in Figure \ref{fig:results-base-individual}.
As before, the thick orange line is the regret curve of the EMaliboo algorithm, while the thin orange lines represent the regret curves of the 10 individual ensemble members.
By definition, at any point in time, the simple regret of the ensemble is equal to the current best (smallest) simple regret among the agents.
However, we often observe significant variance in the performance of the individual agents, as showcased by Figure \ref{fig:results-base-individual}, and it is not uncommon for the leading agent to change over time.
For these reasons, the performance of EMaliboo may be inconsistent at times despite its generally superior performance compared to the centralized PAMaliboo.
The latter method may be preferable for those who favor consistency of the optimization process.

\subsection{Error injection experiments}  \label{subsec:exper-err-inj}
We also conduct additional simulated experiments to observe the change in the behavior of our algorithms in the presence of large prediction errors in the ML model $\hat{g}(\cdot)$.
The performance of our optimization algorithms depends heavily on the ability of the ML models in the acquisition function to predict the constrained resource.
However, it may not always be possible to choose an accurate model.
For instance, prior information about the application to optimize may be too scarce for an appropriate preliminary analysis; alternatively, the chosen ML model may underperform in the initial phase of the optimization, which is a crucial time window for the exploration phase.
We, therefore, create a scenario in which ML predictions are affected by significant errors.

To do so, we inject an error into the model predictions.
Specifically, we multiply the ML prediction by a factor that starts from 1.5 and linearly decreases in the first 50 iterations until it becomes 1 (therefore not affecting the ML prediction anymore after that point).
Formally speaking, at iteration $i$, given $\varepsilon_0 = 1.5$ and $N_{err} = 50$, we change the prediction of the ML model $\hat{g}(\cdot)$ as follows:
\begin{equation}
    \hat{g}_{err}(x) = \hat{g}(x) \, \cdot \, \varepsilon_i \qquad \text{with } \varepsilon_i = \varepsilon_0 - \frac{(\varepsilon_0 - 1) \cdot \min(i, N_{err})}{N_{err}}.
\end{equation}
Note that this change does not apply to OpenTuner since it uses no ML performance model.
Therefore, we compare the new error-injected experiments with the OpenTuner experiments from the previous section.

\begin{figure}[t]
    \centering
    \newcommand\mult{0.45}
    \begin{subfigure}[b]{\mult\linewidth}
        \centering
        \includegraphics[width=\linewidth]{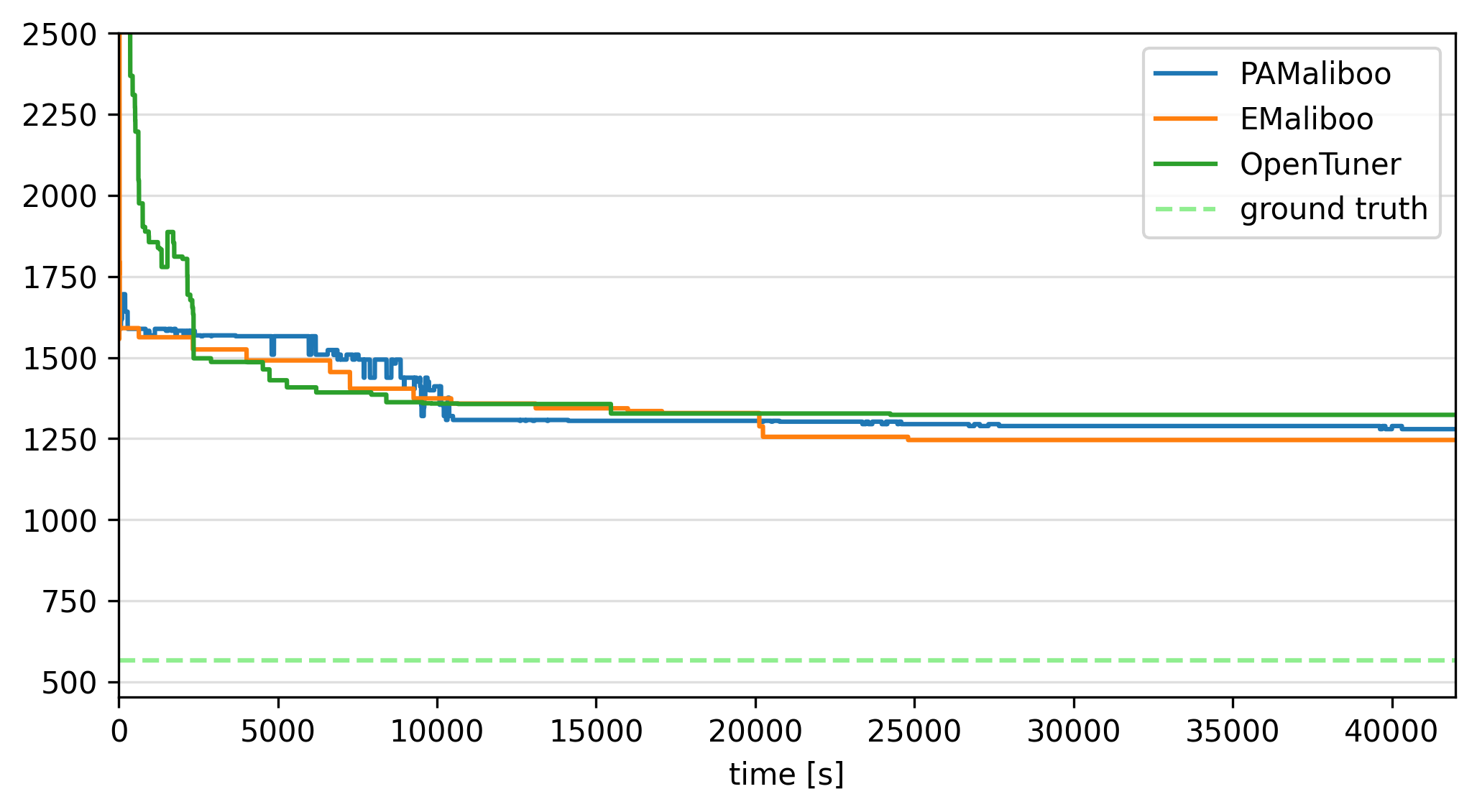}
        \caption{Current best feasible value of the objective function over time.}
        \label{subfig:results-err-inj-a}
    \end{subfigure}
    \begin{subfigure}[b]{\mult\linewidth}
        \centering
        \includegraphics[width=\linewidth]{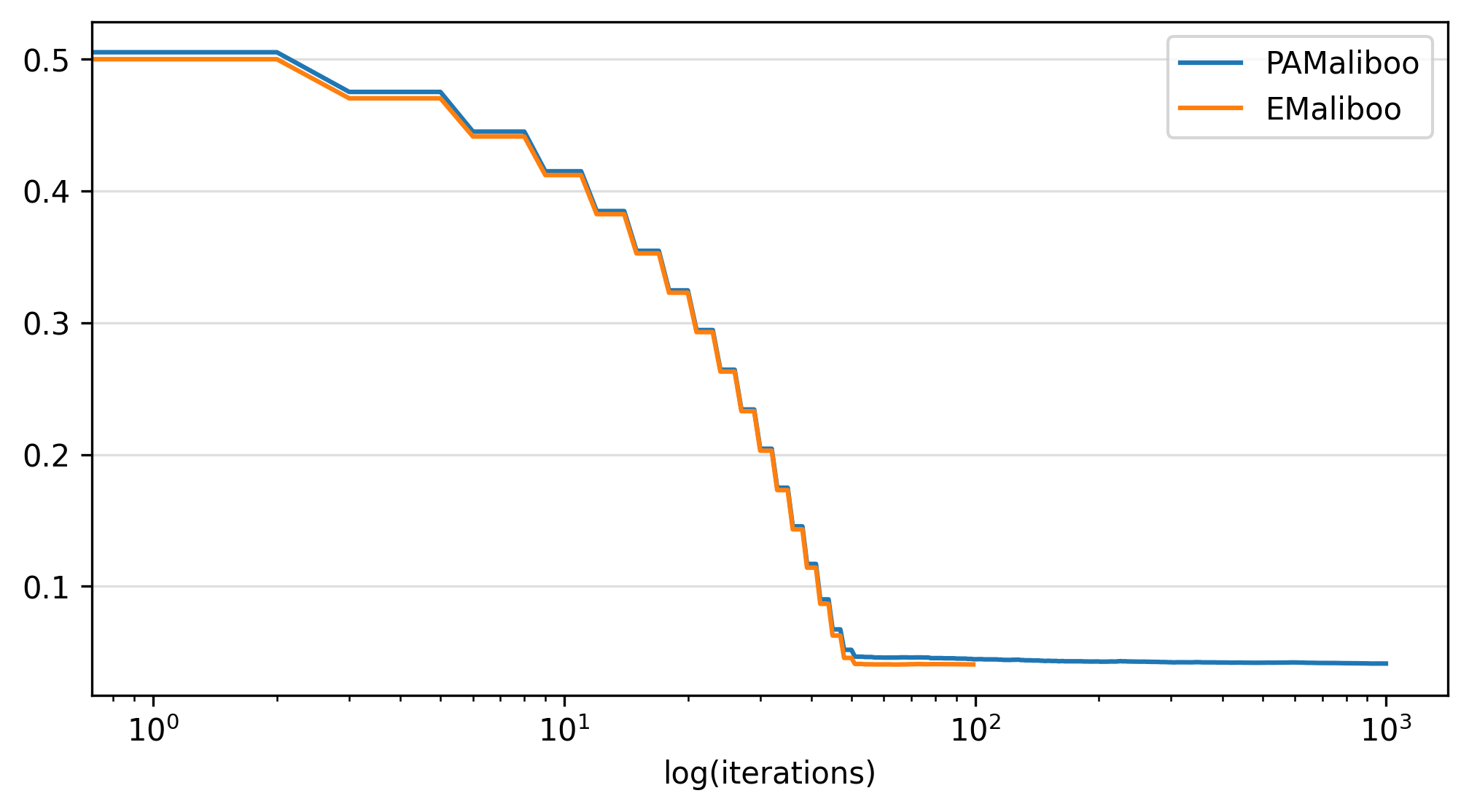}
        \caption{MAPE of ML models over log-scale of iterations.}
        \label{subfig:results-err-inj-b}
    \end{subfigure}
    \begin{subfigure}[b]{\mult\linewidth}
        \centering
        \includegraphics[width=\linewidth]{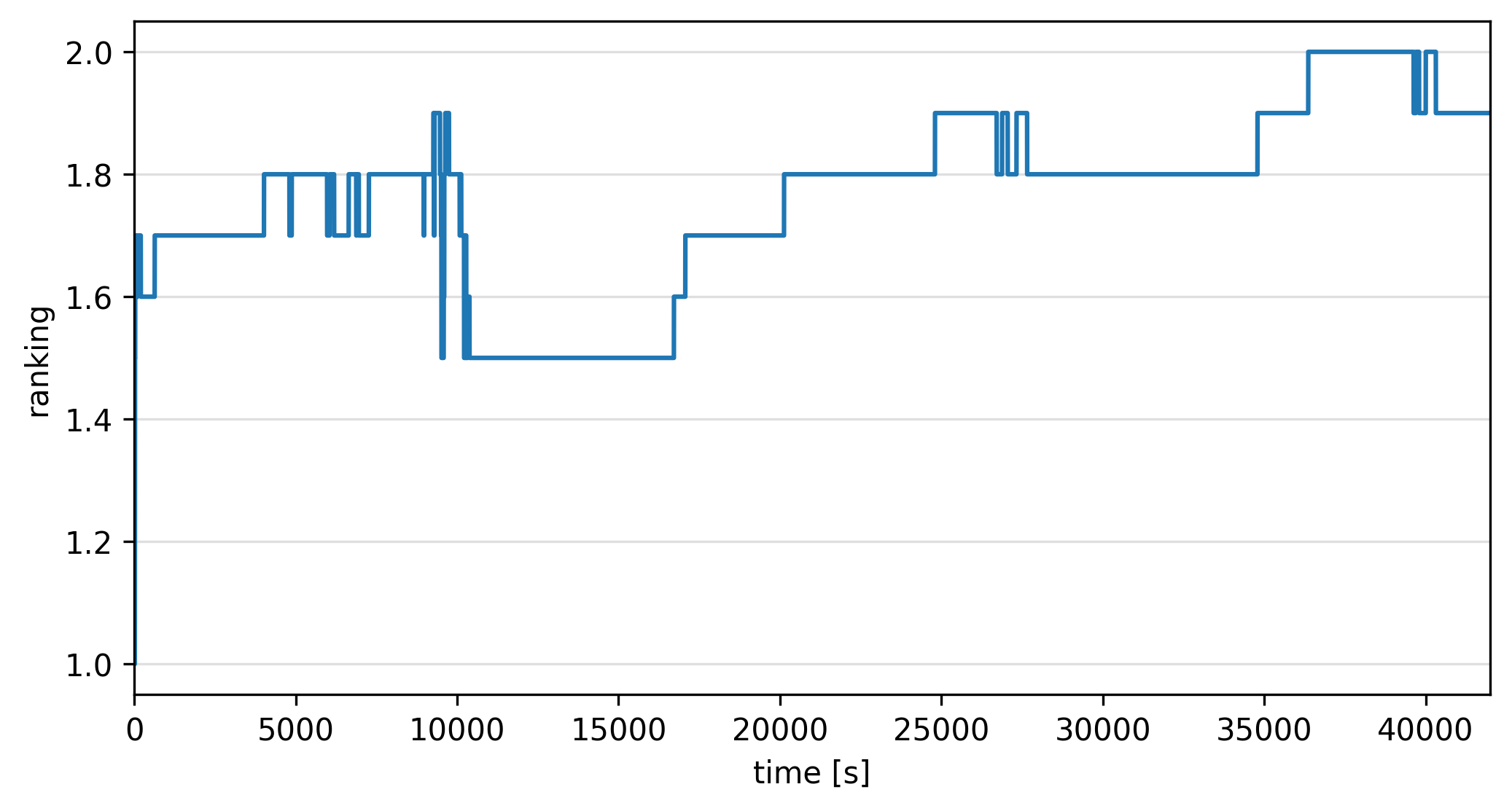}
        \caption{Ranking of PAMaliboo agent vs EMaliboo members over time.}
        \label{subfig:results-err-inj-c}
    \end{subfigure}
    \caption{Experiments with error injection: 10-experiment average of relevant quantities.}
    \label{fig:results-err-inj}
\end{figure}

We conduct these experiments with the same hardware environment and settings as in the previous section.
We show average results across 10 experiments in Figure \ref{fig:results-err-inj}.
In the initial phase, we observe a slowdown in the feasible regret curve of both BO-based techniques (see Figure \ref{subfig:results-err-inj-a}), likely due to a decreased ability of the algorithms to find feasible configurations.
However, after the transitory phase with the injected error (see the initial slope in Figure \ref{subfig:results-err-inj-b}), the models have gathered enough information to converge to about the same regret value as the regular experiment.
At the same time, they still score a better result than the OpenTuner framework in terms of objective function values.
Between the two BO techniques, EMaliboo is the most affected by the injected error.
The likely reason is that the transitory phase affects a more significant portion of the iterations, i.e., the first 50 out of 100 for all agents, namely half of all the iterations.
In the rankings plot of Figure \ref{subfig:results-err-inj-c}, we observe a similar pattern as the previous experiment.
However, the ranking of the centralized model is even closer to 1, confirming the competitiveness of the PAMaliboo approach in the presence of less accurate models.

\subsection{Prototype environment experiments}  \label{subsec:exper-real}
Finally, we run experiments with actual LiGen executions in a real prototype environment.
In this case, we only perform 5 runs of the algorithms (each with a different random seed, as before) due to time and budget constraints.

The environment in which we perform these experiments has a login node with a virtual processor @ 2.2 GHz, 32 GB RAM, 16 cores, and one thread per core; the computation nodes have virtual processors @ 2.2 GHz with 32 GB RAM, 8 cores, and one thread per core.
We use 10 nodes to achieve the same parallelism level as the previous experiments.

\begin{figure}[t]
    \centering
    \newcommand\mult{0.45}
    \begin{subfigure}[b]{\mult\linewidth}
        \centering
        \includegraphics[width=\linewidth]{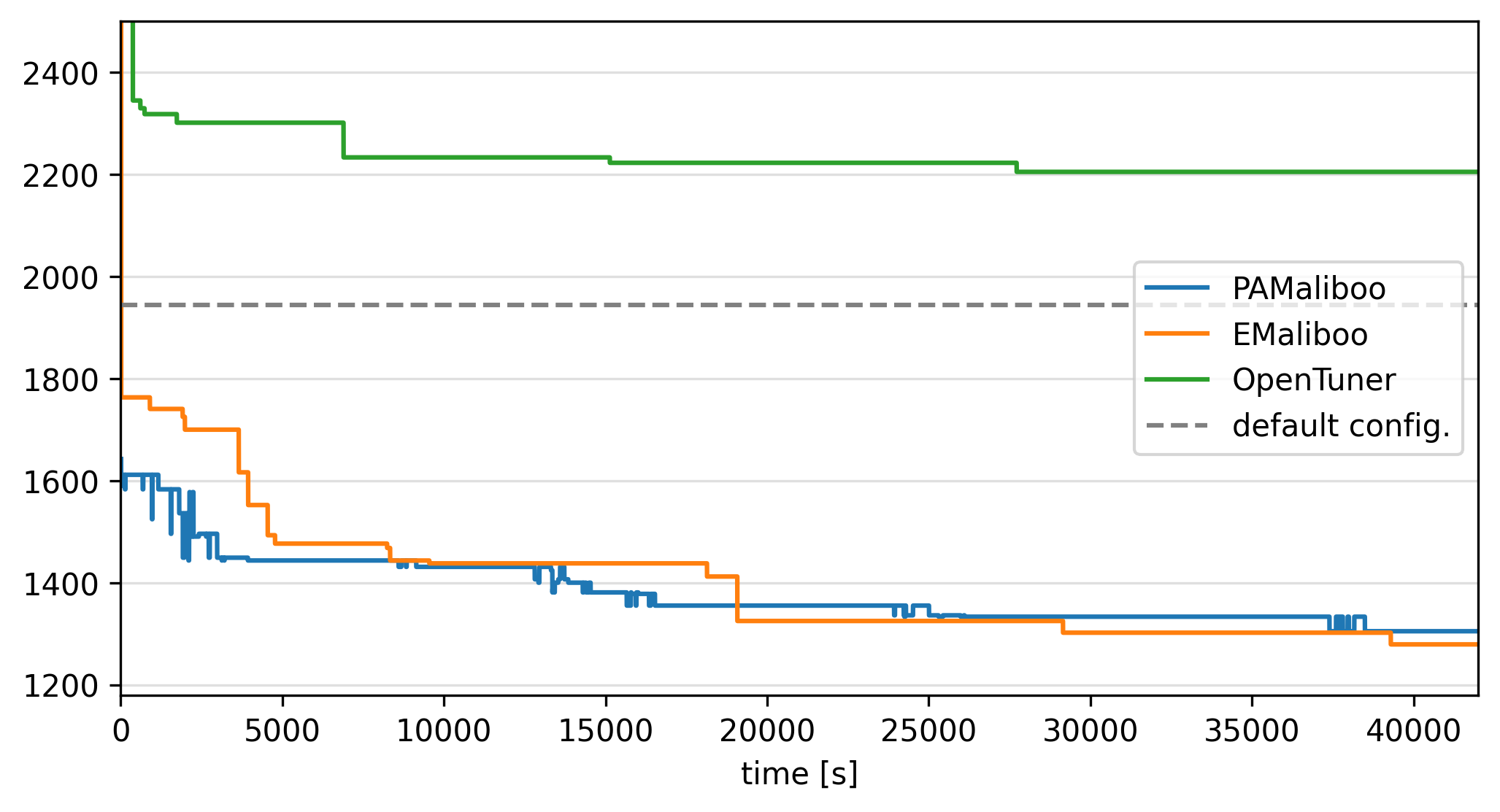}
        \caption{Current best feasible value of the objective function over time.}
        \label{subfig:results-real-a}
    \end{subfigure}
    \begin{subfigure}[b]{\mult\linewidth}
        \centering
        \includegraphics[width=\linewidth]{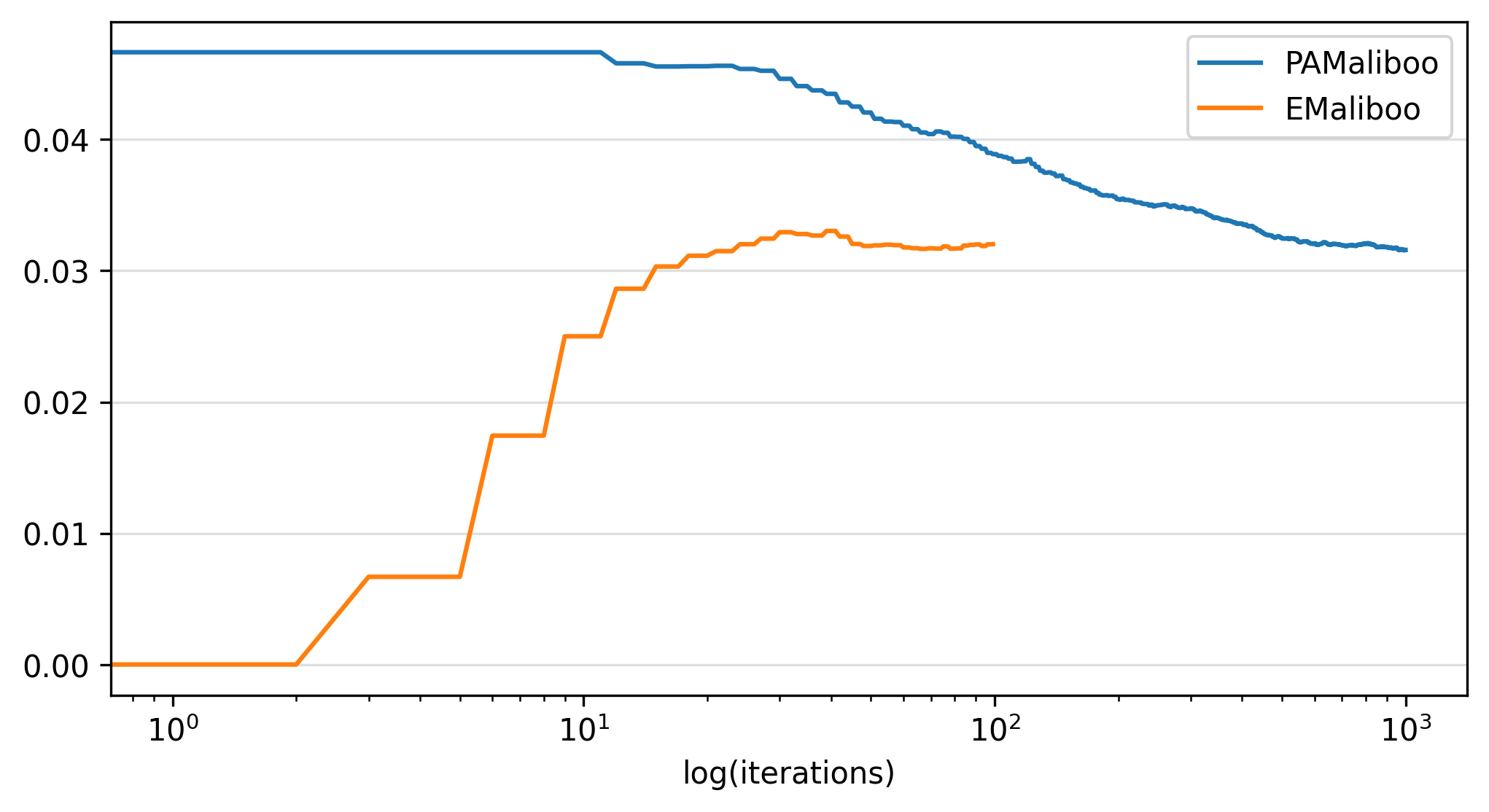}
        \caption{MAPE of ML models over log-scale of iterations.}
        \label{subfig:results-real-b}
    \end{subfigure}
    \begin{subfigure}[b]{\mult\linewidth}
        \centering
        \includegraphics[width=\linewidth]{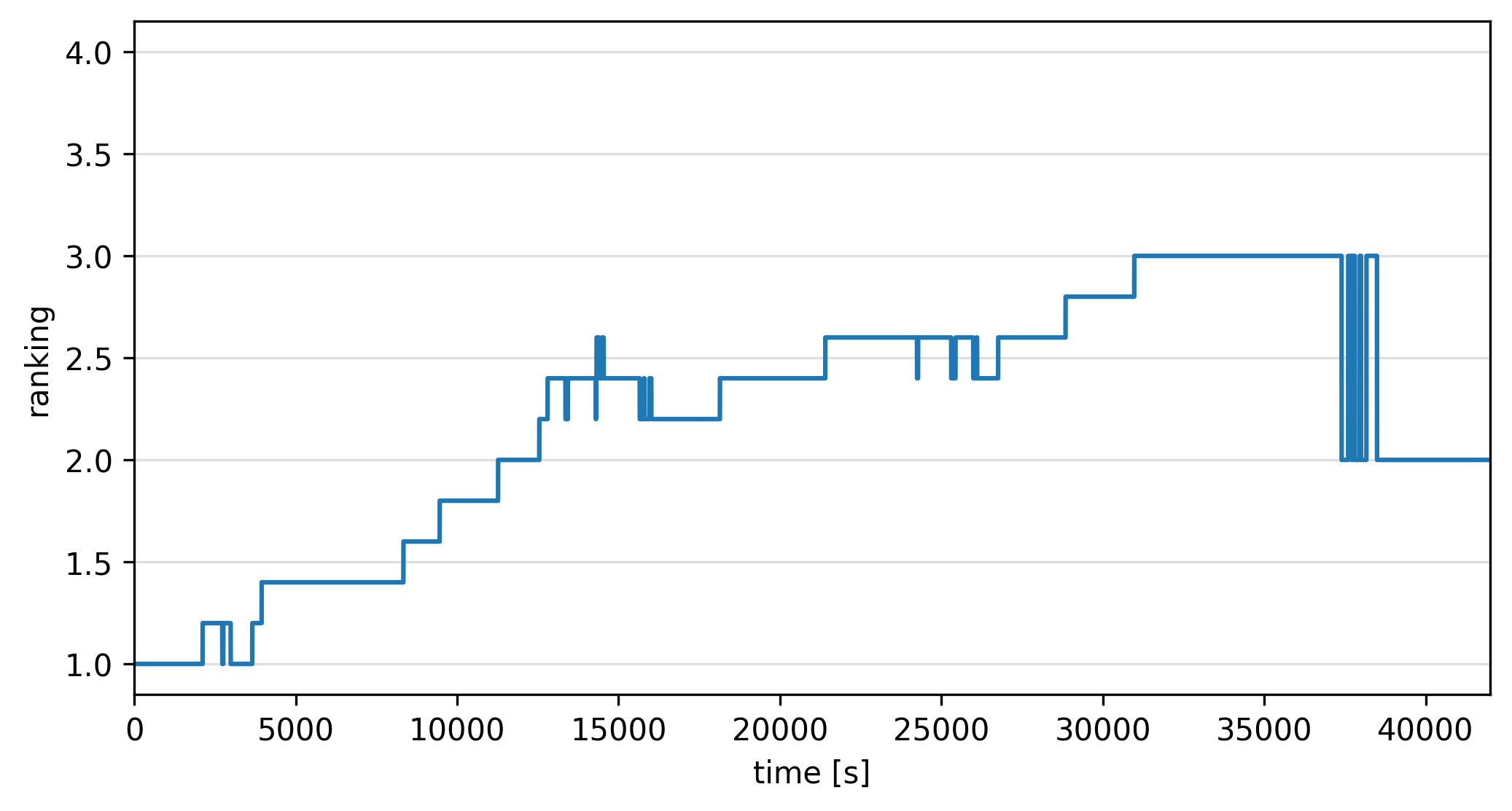}
        \caption{Ranking of PAMaliboo agent vs EMaliboo members over time.}
        \label{subfig:results-real-c}
    \end{subfigure}
    \caption{Prototype environment experiments: 5-experiment average of relevant quantities.}
    \label{fig:results-real}
\end{figure}
We report the average results across the 5 runs in Figure \ref{fig:results-real}.
As shown in Figure \ref{subfig:results-real-a}, the difference in simple regret between OpenTuner (in green) and the BO-based techniques is even more pronounced than in previous experiments.
The ensemble-based EMaliboo (in orange) can quickly catch up to the centralized PAMaliboo (in blue) despite an initial slowdown, thanks to its excellent exploration capabilities.
Meanwhile, OpenTuner is stuck to suboptimal configurations, achieving a final objective function value about 72\% larger than EMaliboo.

The ground truth for this experiment is not known since it would require an exhaustive search of the entire optimization domain.
Instead, we compare the performance of the algorithm against the initial default configuration of LiGen, chosen according to experts' knowledge, without using autotuning techniques.
This configuration corresponds to the horizontal dashed grey line in Figure \ref{subfig:results-real-a}.
The best configurations found by both BO-based techniques achieve a value up to 35\% smaller than the value corresponding to the default configuration.

As evidenced by Figures and \ref{subfig:results-real-a} and \ref{subfig:results-real-c}, EMaliboo outperforms the other algorithms.
Also, the prediction errors in Figure \ref{subfig:results-real-b} have a similar trend as the previous cases, settling on values around 3\%.
Overall, the experiments performed in the prototype environment confirm the improvement of BO-based techniques compared to the state-of-the-art OpenTuner.
Our techniques also bring a significant improvement compared to the expert-chosen default configuration.

\subsection{Discussion}  \label{subsubsec:exper-discussion}
In Table \ref{tab:incumbents-base}, we report the average incumbents, i.e., the smallest objective function values observed, for each technique at the end of the observed time window for all types of experiments.
\begin{table}
    \centering
    \begin{tabular}{|c|c|c|c|}
        \hline
        \textbf{Technique} & \textbf{Simulated} & \textbf{Error inj.} & \textbf{Prototype env.} \\
        \hline
        EMaliboo              & 1154.6 & 1245.6 & 1279.1 \\
        PAMaliboo             & 1273.6 & 1279.2 & 1304.9 \\
        OpenTuner             & 1323.2 & N/A    & 2204.9 \\
        default configuration & N/A    & N/A    & 1943.9 \\
        \hline
    \end{tabular}
    \caption{Average final incumbents values for all experiments}
    \label{tab:incumbents-base}
\end{table}
On average, EMaliboo achieves a 13\% improvement compared to OpenTuner in simulated experiments.
In the error injection experiments, the performance of PAMaliboo is barely affected, while EMaliboo has a more noticeable increase of about 7\% compared to the simulated experiment.
In the prototype environment experiments, the improvement of EMaliboo and PAMaliboo compared to OpenTuner further increases to 40\% and 42\%, respectively.
Also, EMaliboo and PAMaliboo improve on the expert-chosen default configuration by 35\% and 33\%, respectively, while OpenTuner is at best 13\% worse than the default.

In conclusion, our proposed BO-based techniques outperform the state-of-the-art OpenTuner algorithm in all experiments considered.
EMaliboo is the overall superior technique in exploring capabilities and simple regret achieved, provided that the ML models perform well in predicting the constrained resource (as in our case, according to our preliminary analysis).
On the other hand, PAMaliboo is the more robust choice because it collects a more significant amount of data within the centralized models (both the GP for the objective function and the ML model to estimate the constrained resource).
It is best suited for scenarios characterized by large prediction errors in which the user seeks a more conservative approach in the search for the optimal configuration.

\section{Related work}  \label{sec:related}
This section reviews various approaches to adapting applications for optimal performance in HPC.
We categorize existing works into specific and generic autotuners, highlighting the evolution from manual profiling to automated and dynamic tuning methods (Section \ref{subsec:related-autotuning-hpc}).
We then shift the focus to the application of BO-based approaches to the HPC setting (Section \ref{subsec:related-bo-hpc}).

\subsection{Autotuning in HPC}  \label{subsec:related-autotuning-hpc}
While several fields of computer science aim to ease the differences in computer architectures and hardware capabilities, to get the most performance out of our applications, users must adapt them to the execution environment.
Adaptations can happen at different phases of the application deployment and can affect mostly extra-functional properties of executions like time to result, accuracy, and power consumption.

In the HPC scenario, adaptivity plays a pivotal role in ensuring the optimal usage of the powerful resources available \cite{balaprakash2018autotuning, gocht2019q}.
As we aim towards maximum performance, it becomes mandatory to carefully select the best values of all the tunable knobs provided by the application.
Users tend to perform this selection in an iterative approach, with a sequence of choices and evaluations, i.e., \textit{application profiling} \cite{galimberti2023oscar}.
The issues with application profiling arise when the number of possible configurations becomes too large to explore, thus introducing the need for automation and efficient strategies through autotuners.
This aspect is fundamental since evaluating a configuration requires time and computational resources, thus affecting the overall cost of the computation.
Autotuners carefully choose which configurations to test, often through iterative algorithms, trying to model how the application reacts to the setting.
When the built model is accurate, they use it to predict the best configuration, thus avoiding testing them all \cite{nugteren2015cltune, ansel2014opentuner}.
Nonetheless, choosing the best model to represent the application reaction and selecting the proper configurations to test is a non-trivial problem, and the state-of-the-art features many alternative approaches.

Autotuners diversify depending on whether they focus on a specific type of application or propose a domain-agnostic generic approach.
In the first category, we can find works like CLTune \cite{nugteren2015cltune}, GLINDA \cite{Shen:2013:GFA:2482767.2482785}, and Sepya \cite{kamil2012productive}.
The first two efforts target OpenCL applications, while the third focuses on stencil computations.
Alongside these works, we can also consider QuickStep \cite{misailovic2013parallelizing}, Paraprox \cite{samadi2014paraprox}, and PowerGAUGE \cite{dorn2017automatically}, which attempt to introduce adaptation properties in applications by exposing software knobs.
These works enable the possibility of the accuracy/throughput trade-off, making them application-independent but specific to a single optimization setting.

In the second category, we can consider more generic efforts, i.e., not tailored to a single application characteristic.
Works like OpenTuner \cite{ansel2014opentuner} and  Auto-Tuning Framework (ATF) \cite{rasch2018atf} belong to this category since their authors designed them to explore an ample configuration space and efficiently find the best one.
In OpenTuner \cite{ansel2014opentuner}, a meta-search algorithm guides the exploration process, allocating more tests to techniques that perform well.
Its default meta-algorithm is the ``multi-armed bandit with sliding window, area under the curve credit assignment,'' also known as AUC Bandit.
The techniques coordinated by the meta-algorithm include classical optimization methods such as differential evolution and greedy mutation variants.
Individual techniques share results through a common database so that improvements made by one of them can also benefit the others.
On the other hand, ATF \cite{rasch2018atf} is a generic tool that is agnostic to the programming language used, application domain, and tuning objective.
It allows interdependencies between tuning parameters by introducing parameter constraints and allows large parameter ranges by optimizing the process of search space generation.

Alongside those works, we must also consider Petabricks \cite{ding2015autotuning} and Capri \cite{sui2016proactive}, which can infer the effect of input features on the choice of the best configuration.
In this paper, we developed solutions that belong to this category since we are interested in discovering the best configuration of the LiGen application during its deployment before the production phase.

While all these works performed profiling and adaptation at application design time, others showed the possibility of doing these steps at runtime.
We refer to those frameworks as dynamic autotuners, as they can learn and enact the best configuration during application execution, effectively reducing the deployment time.
Efforts like \cite{laurenzano2016input, miguel2016anytime} are good examples of dynamic autotuners, as they can model the response of the system to various configurations during its execution, and they can react to changes as they happen.
While powerful, these tools fit more with a mutable scenario and thus are out of the scope of this analysis.

In \cite{gadioli2021tunable}, the authors introduce tunable approximations to a mini-app derived by the LiGen software.
By exposing five software knobs, they attempt to find the optimal performance-accuracy trade-off by estimating the time-to-solution of the mini-app via linear regression.
Their work shows the advantages of a tuning phase for the optimization of LiGen.
However, the configuration space was fully analyzed, and no effort toward an efficient exploration was made.


\subsection{Bayesian Optimization in HPC}  \label{subsec:related-bo-hpc}
BO is a popular and successful choice in the literature for design space exploration, autotuning, and optimization in HPC settings.
However, most of the existing efforts do not support parallel exploration.
Authors of \cite{fu2024design} leverage multi-objective BO to find the best Power, Performance, and Area (PPA) configurations for HPC multiprocessor system architectures.
They use virtual prototyping tools to simulate objective metrics and Random Forest priors to model them, creating and updating a Pareto front of optimal solutions.
In Section \ref{subsec:backgr-bo-algos}, we mentioned \cite{egele2022asynchronous}, an ensemble-based parallel approach deployed on multiple HPC nodes.
Each node runs its own independent BO instance with a Random Forest Regressor prior and the Upper Confidence Bound (UCB) acquisition function, each with a different value for the UCB hyperparameter.
Individual agents communicate their results to each other asynchronously so that they will be part of the next posterior distribution update.
HiPerBOt \cite{menon2020auto} is an active learning framework based on Expected Improvement (EI) BO and dedicated to tuning HPC applications.
It also supports transfer learning to recycle knowledge from low-cost source data to more complex scenarios when limited resources are available for data collection at scale.
Authors of \cite{miyazaki2018bayesian} minimize the energy efficiency of a GPU cluster system via BO with the Gaussian Process Mutual Information (GP-MI) acquisition function, an enhancement of the traditional UCB acquisition \cite{contal2014gaussian}.
Such an approach enabled them to score second in the Green500 list, a popular energy-efficiency ranking of supercomputers.
Finally, we mention the original MALIBOO framework \cite{guindani2024integrating}, which is the basis of this work.
MALIBOO integrates ML models to estimate information about constrained resources into the BO mechanism.
However, it is a sequential optimization technique which makes it unsuitable for the scope of this work, as it cannot fully harness the computational power of HPC settings, reducing the potential advantage provided by the efficient configuration selection.

\section{Conclusions}  \label{sec:conclusion}
Efficient tuning of highly parameterized applications is a critical challenge in HPC.
Often, one must deal with the enormous configuration space of an application and with the opportunity to use highly parallel environments and resources to perform the exploration.
In this work, we have analyzed the problem by applying an efficient strategy using two parallel variants of an autotuning technique to LiGen, a real-world virtual screening software for drug discovery.
Navigating the throughput-accuracy trade-off for the optimal configuration of LiGen is particularly crucial due to the sheer size of its configuration space and the influence of the input software knobs on the efficiency of the screening process.

The two autotuning variants are designed for constrained optimization in distributed HPC environments.
The ensemble-based method EMaliboo and the pure asynchronous method PAMaliboo take different routes in the search for the best application configuration but are both based on BO.
They also integrate predictions from ML models to estimate relevant quantities subject to constraints. They both can execute the application to optimize in a unified framework meant for HPC systems, exploiting either a centralized or distributed deployment.

In the experimental validation campaign, we found that the two proposed techniques serve different purposes.
EMaliboo is the overall superior technique in exploring capabilities, provided that the ML models perform well in predicting the constrained resource.
On the other hand, PAMaliboo is the more robust and conservative choice since it trains models with larger centralized datasets and is best suited for scenarios characterized by inaccurate ML predictions.
Overall, both techniques outperform the state-of-the-art OpenTuner autotuning framework and the default LiGen configuration chosen by experts by upwards of 35--42\% in terms of simple regret.


\section*{Acknowledgments}
The European Commission has partially funded this work under the Horizon 2020 Grant Agreement number 956137 LIGATE: LIgand Generator and portable drug discovery platform AT Exascale, as part of the European High-Performance Computing (EuroHPC) Joint Undertaking program.

\bibliographystyle{elsarticle-num}
\bibliography{bibliography}

\begin{thebibliography}{10}
\expandafter\ifx\csname url\endcsname\relax
  \def\url#1{\texttt{#1}}\fi
\expandafter\ifx\csname urlprefix\endcsname\relax\def\urlprefix{URL }\fi
\expandafter\ifx\csname href\endcsname\relax
  \def\href#1#2{#2} \def\path#1{#1}\fi

\bibitem{polishchuk2013estimation}
P.~G. Polishchuk, T.~I. Madzhidov, A.~Varnek, Estimation of the size of drug-like chemical space based on gdb-17 data, Journal of computer-aided molecular design 27 (2013) 675--679.

\bibitem{matter2011application}
H.~Matter, C.~Sotriffer, Applications and Success Stories in Virtual Screening, John Wiley \& Sons, Ltd, 2011, Ch.~12, pp. 319--358.

\bibitem{allegretti2022repurposing}
M.~Allegretti, M.~C. Cesta, M.~Zippoli, A.~Beccari, C.~Talarico, F.~Mantelli, E.~M. Bucci, L.~Scorzolini, E.~Nicastri, Repurposing the estrogen receptor modulator raloxifene to treat sars-cov-2 infection, Cell Death \& Differentiation 29~(1) (2022) 156--166.

\bibitem{murugan2022review}
N.~A. Murugan, A.~Podobas, D.~Gadioli, E.~Vitali, G.~Palermo, S.~Markidis, A review on parallel virtual screening softwares for high-performance computers, Pharmaceuticals 15~(1) (2022) 63.

\bibitem{pagadala2017software}
N.~S. Pagadala, K.~Syed, J.~Tuszynski, Software for molecular docking: a review, Biophysical reviews 9~(2) (2017) 91--102.

\bibitem{su2018comparative}
M.~Su, Q.~Yang, Y.~Du, G.~Feng, Z.~Liu, Y.~Li, R.~Wang, Comparative assessment of scoring functions: the casf-2016 update, Journal of chemical information and modeling 59~(2) (2018) 895--913.

\bibitem{9309041}
J.~Sewall, S.~J. Pennycook, D.~Jacobsen, T.~Deakin, S.~McIntosh-Smith, Interpreting and visualizing performance portability metrics, in: 2020 IEEE/ACM International Workshop on Performance, Portability and Productivity in HPC (P3HPC), 2020, pp. 14--24.
\newblock \href {http://dx.doi.org/10.1109/P3HPC51967.2020.00007} {\path{doi:10.1109/P3HPC51967.2020.00007}}.

\bibitem{gadioli2021tunable}
D.~Gadioli, G.~Palermo, S.~Cherubin, E.~Vitali, G.~Agosta, C.~Manelfi, A.~R. Beccari, C.~Cavazzoni, N.~Sanna, C.~Silvano, Tunable approximations to control time-to-solution in an hpc molecular docking mini-app, The Journal of Supercomputing 77 (2021) 841--869.

\bibitem{verma2015large}
A.~Verma, L.~Pedrosa, M.~Korupolu, D.~Oppenheimer, E.~Tune, J.~Wilkes, Large-scale cluster management at google with borg, in: Proceedings of the tenth european conference on computer systems, 2015, pp. 1--17.

\bibitem{gocht2019q}
A.~Gocht, R.~Sch{\"o}ne, M.~Bielert, Q-learning inspired self-tuning for energy efficiency in hpc, in: 2019 International Conference on High Performance Computing \& Simulation (HPCS), IEEE, 2019, pp. 344--347.

\bibitem{balaprakash2018autotuning}
P.~Balaprakash, J.~Dongarra, T.~Gamblin, M.~Hall, J.~K. Hollingsworth, B.~Norris, R.~Vuduc, Autotuning in high-performance computing applications, Proceedings of the IEEE 106~(11) (2018) 2068--2083.

\bibitem{rasch2018atf}
A.~Rasch, S.~Gorlatch, Atf: A generic auto-tuning framework, in: Proceedings of the 27th International Symposium on High-Performance Parallel and Distributed Computing, 2018, pp. 3--4.

\bibitem{dorn2017automatically}
J.~Dorn, J.~Lacomis, W.~Weimer, S.~Forrest, Automatically exploring tradeoffs between software output fidelity and energy costs, IEEE Transactions on Software Engineering.

\bibitem{frazier2018tutorial}
P.~I. Frazier, {A Tutorial on Bayesian Optimization}, arXiv preprint arXiv:1807.02811.

\bibitem{guindani2024integrating}
B.~Guindani, D.~Ardagna, A.~Guglielmi, R.~Rocco, G.~Palermo, {Integrating Bayesian Optimization and Machine Learning for the Optimal Configuration of Cloud Systems}, IEEE Transactions on Cloud Computing 12~(1) (2024) 277--294.

\bibitem{Beranek2024HQ}
J.~Ber{\'a}nek, A.~B{\"o}hm, G.~Palermo, J.~Martinovi{\v c}, B.~Jans{\'\i}k, Hyperqueue: Efficient and ergonomic task graphs on hpc clusters, SoftwareX 27.
\newblock \href {http://dx.doi.org/https://doi.org/10.1016/j.softx.2024.101814} {\path{doi:https://doi.org/10.1016/j.softx.2024.101814}}.

\bibitem{9817028}
D.~Gadioli, E.~Vitali, F.~Ficarelli, C.~Latini, C.~Manelfi, C.~Talarico, C.~Silvano, C.~Cavazzoni, G.~Palermo, A.~R. Beccari, Exscalate: An extreme-scale virtual screening platform for drug discovery targeting polypharmacology to fight sars-cov-2, IEEE Transactions on Emerging Topics in Computing 11~(1) (2023) 170--181.
\newblock \href {http://dx.doi.org/10.1109/TETC.2022.3187134} {\path{doi:10.1109/TETC.2022.3187134}}.

\bibitem{ansel2014opentuner}
J.~Ansel, S.~Kamil, K.~Veeramachaneni, J.~Ragan-Kelley, J.~Bosboom, U.-M. O'Reilly, S.~Amarasinghe, Opentuner: An extensible framework for program autotuning, in: Proceedings of the 23rd international conference on Parallel architectures and compilation, 2014, pp. 303--316.

\bibitem{egele2022asynchronous}
R.~Egele, J.~Gouneau, V.~Vishwanath, I.~Guyon, P.~Balaprakash, Asynchronous distributed bayesian optimization at hpc scale, arXiv preprint arXiv:2207.00479.

\bibitem{frisby2021asynchronous}
T.~S. Frisby, Z.~Gong, C.~J. Langmead, Asynchronous parallel bayesian optimization for ai-driven cloud laboratories, Bioinformatics 37~(Supplement 1) (2021) i451--i459.

\bibitem{kandasamy2018parallelised}
K.~Kandasamy, A.~Krishnamurthy, J.~Schneider, B.~P{\'o}czos, Parallelised bayesian optimisation via thompson sampling, in: International Conference on Artificial Intelligence and Statistics, PMLR, 2018, pp. 133--142.

\bibitem{snoek2012practical}
J.~Snoek, H.~Larochelle, R.~P. Adams, Practical bayesian optimization of machine learning algorithms, Advances in Neural Information Processing Systems 25 (2012) 2951--2959.

\bibitem{schonlau1998global}
M.~Schonlau, W.~J. Welch, D.~R. Jones, {Global versus Local Search in Constrained Optimization of Computer Models}, IMS Lecture Notes-Monograph Series (1998) 11--25.

\bibitem{rasmussen2006gaussian}
C.~E. Rasmussen, C.~K. Williams, et~al., Gaussian processes for machine learning, Vol.~1, Springer, 2006.

\bibitem{ginsbourger2010kriging}
D.~Ginsbourger, R.~Le~Riche, L.~Carraro, Kriging is well-suited to parallelize optimization, Computational intelligence in expensive optimization problems (2010) 131--162.

\bibitem{assran2020advances}
M.~Assran, A.~Aytekin, H.~R. Feyzmahdavian, M.~Johansson, M.~G. Rabbat, Advances in asynchronous parallel and distributed optimization, Proceedings of the IEEE 108~(11) (2020) 2013--2031.

\bibitem{zhang2020efficient}
S.~Zhang, F.~Yang, D.~Zhou, X.~Zeng, An efficient asynchronous batch bayesian optimization approach for analog circuit synthesis, in: 2020 57th ACM/IEEE Design Automation Conference (DAC), IEEE, 2020, pp. 1--6.

\bibitem{ginsbourger2007multi}
D.~Ginsbourger, R.~Le~Riche, L.~Carraro, A multi-points criterion for deterministic parallel global optimization based on kriging, in: International Conference on Nonconvex Programming (NCP07), 2007, pp. 1--30.

\bibitem{kawaguchi2015bayesian}
K.~Kawaguchi, L.~P. Kaelbling, T.~Lozano-P{\'e}rez, Bayesian optimization with exponential convergence, Advances in Neural Information Processing Systems 28 (2015) 2809--2817.

\bibitem{kutzner2015best}
C.~Kutzner, S.~P{\'a}ll, M.~Fechner, A.~Esztermann, B.~L. de~Groot, H.~Grubm{\"u}ller, Best bang for your buck: Gpu nodes for gromacs biomolecular simulations (2015).

\bibitem{10.1145/3235830.3235835}
E.~Vitali, D.~Gadioli, G.~Palermo, A.~Beccari, C.~Silvano, Accelerating a geometric approach to molecular docking with openacc, in: Proceedings of the 6th International Workshop on Parallelism in Bioinformatics, PBio 2018, Association for Computing Machinery, 2018, p. 45–51.
\newblock \href {http://dx.doi.org/10.1145/3235830.3235835} {\path{doi:10.1145/3235830.3235835}}.

\bibitem{vitali2022gpuoptimized}
E.~Vitali, F.~Ficarelli, M.~Bisson, D.~Gadioli, G.~Accordi, M.~Fatica, A.~R. Beccari, G.~Palermo, Gpu-optimized approaches to molecular docking-based virtual screening in drug discovery: A comparative analysis, Journal of Parallel and Distributed Computing 186.
\newblock \href {http://dx.doi.org/https://doi.org/10.1016/j.jpdc.2023.104819} {\path{doi:https://doi.org/10.1016/j.jpdc.2023.104819}}.

\bibitem{9651263}
S.~Markidis, D.~Gadioli, E.~Vitali, G.~Palermo, Understanding the i/o impact on the performance of high-throughput molecular docking, in: 2021 IEEE/ACM Sixth International Parallel Data Systems Workshop (PDSW), 2021, pp. 9--14.
\newblock \href {http://dx.doi.org/10.1109/PDSW54622.2021.00007} {\path{doi:10.1109/PDSW54622.2021.00007}}.

\bibitem{guindani2023amllibrary}
B.~Guindani, M.~Lattuada, D.~Ardagna, {AMLLibrary: An AutoML Approach for Performance Prediction}, in: 37th International Conference on Modelling and Simulation (ECMS), Vol.~37, ECMS, 2023, pp. 241--247.

\bibitem{nelder1965simplex}
J.~A. Nelder, R.~Mead, A simplex method for function minimization, The computer journal 7~(4) (1965) 308--313.

\bibitem{galimberti2023oscar}
E.~Galimberti, B.~Guindani, F.~Filippini, H.~Sedghani, D.~Ardagna, S.~Risco, G.~Molt{\'o}, M.~Caballer, {OSCAR-P and aMLLibrary: Performance Profiling and Prediction of Computing Continua Application}, in: 1st Workshop on Artificial Intelligence for Performance Modeling, Prediction, and Control (AIPerf), ACM, 2023, pp. 139--146.

\bibitem{nugteren2015cltune}
C.~Nugteren, V.~Codreanu, Cltune: A generic auto-tuner for opencl kernels, in: Embedded Multicore/Many-core Systems-on-Chip (MCSoC), 2015 IEEE 9th International Symposium on, IEEE, 2015, pp. 195--202.

\bibitem{Shen:2013:GFA:2482767.2482785}
J.~Shen, A.~L. Varbanescu, H.~Sips, M.~Arntzen, D.~G. Simons, Glinda: A framework for accelerating imbalanced applications on heterogeneous platforms, in: Proceedings of the ACM International Conference on Computing Frontiers, CF '13, ACM, 2013, pp. 14:1--14:10.
\newblock \href {http://dx.doi.org/10.1145/2482767.2482785} {\path{doi:10.1145/2482767.2482785}}.

\bibitem{kamil2012productive}
S.~A. Kamil, Productive high performance parallel programming with auto-tuned domain-specific embedded languages, University of California, Berkeley, 2012.

\bibitem{misailovic2013parallelizing}
S.~Misailovic, D.~Kim, M.~Rinard, Parallelizing sequential programs with statistical accuracy tests, ACM Transactions on Embedded Computing Systems (TECS) 12~(2s) (2013) 88.

\bibitem{samadi2014paraprox}
M.~Samadi, D.~A. Jamshidi, J.~Lee, S.~Mahlke, Paraprox: Pattern-based approximation for data parallel applications, ACM SIGPLAN Notices 49~(4) (2014) 35--50.

\bibitem{ding2015autotuning}
Y.~Ding, J.~Ansel, K.~Veeramachaneni, X.~Shen, U.-M. O’Reilly, S.~Amarasinghe, Autotuning algorithmic choice for input sensitivity, ACM SIGPLAN Notices 50~(6) (2015) 379--390.

\bibitem{sui2016proactive}
X.~Sui~et al., Proactive control of approximate programs, ACM SIGOPS Operating Systems Review.

\bibitem{laurenzano2016input}
M.~A. Laurenzano~et al., Input responsiveness: using canary inputs to dynamically steer approximation, ACM SIGPLAN Notices.

\bibitem{miguel2016anytime}
J.~S. Miguel, N.~E. Jerger, The anytime automaton, ACM SIGARCH Computer Architecture News 44~(3) (2016) 545--557.

\bibitem{fu2024design}
V.~Fu, L.~Zaourar, A.~Munier-Kordon, M.~Duranton, Design space exploration of hpc systems with random forest-based bayesian optimization, in: Proceedings of the 16th Workshop on Rapid Simulation and Performance Evaluation for Design, 2024, pp. 9--15.

\bibitem{menon2020auto}
H.~Menon, A.~Bhatele, T.~Gamblin, Auto-tuning parameter choices in hpc applications using bayesian optimization, in: 2020 IEEE International Parallel and Distributed Processing Symposium (IPDPS), IEEE, 2020, pp. 831--840.

\bibitem{miyazaki2018bayesian}
T.~Miyazaki, I.~Sato, N.~Shimizu, Bayesian optimization of hpc systems for energy efficiency, in: High Performance Computing: 33rd International Conference, ISC High Performance 2018, Frankfurt, Germany, June 24-28, 2018, Proceedings 33, Springer, 2018, pp. 44--62.

\bibitem{contal2014gaussian}
E.~Contal, V.~Perchet, N.~Vayatis, Gaussian process optimization with mutual information, in: International Conference on Machine Learning, PMLR, 2014, pp. 253--261.

\end{thebibliography}

\end{document}